\documentclass[final,3p,times,onecolumn]{elsarticle}
\usepackage{amsmath,amssymb,amsfonts,scalerel}
\usepackage{blindtext}
\usepackage{graphicx}
\usepackage{adjustbox}
\usepackage{epsfig}
\usepackage{float}
\usepackage{subfig}
\usepackage{xcolor}
\usepackage[colorlinks=true,citecolor=blue,linkcolor=blue,urlcolor=blue]{hyperref}
\usepackage{setspace}
\usepackage{multirow}
\usepackage{multicol}
\usepackage{url}
\usepackage{siunitx}
\usepackage{hyperref}
\usepackage{rotating}
\usepackage{geometry}
\biboptions{sort,compress}
\journal{Computational Condensed Matter}

\begin{document}

\begin{frontmatter}

\title{First principles and Monte Carlo studies of adsorption 
       and desorption properties from 
       Pd$\rm_{1-x}$Ag$\rm_{{x}}$~surface alloy}
\author{S.S.~Awulachew$^{1}$}
\ead{sshinie21@gmail.com}
\author[]{K.N.~Nigussa$^{2}$\corref{cor1}}
\cortext[cor1]{Corresponding author:\ kenate.nemera@aau.edu.et\ 
               (K.N.~Nigussa)}
               
\address{$^1$Department of Physics, \ Haramaya University,\ P.O. Box 138, \ Dire Dawa,\ Ethiopia\\
         $^2$Department of Physics,\ Addis Ababa University,\ P.O. Box 1176,\ Addis Ababa,\ Ethiopia}
                  
\begin{abstract}
The FCC structure of Pd$\rm_{1-x}$Ag$\rm_{x}$~($\rm{x}=$~0.25,\ 
0.50,\ 0.75)\ alloys is\ considered as\ a fuel cell component\ 
in this study.\ We have\ looked into its\ qualities as a\ 
component of a fuel\ cell to see\ whether it could\ be\ 
potentially\ used\ as\ an\ alternative\ replacement\ of\ 
the\ Pt\ catalyst.\ We used Density\ Functional Theory\ 
(DFT)\ to study H and CO\ interaction with the\ 
surface,\ and Kinetic\ Monte Carlo~(KMC) to\ 
study H and CO\ desorption from the surface.\ 
The bulk modulus\ and equilibrium crystal\ 
structures\ of Pd$\rm_{1-x}$Ag$\rm_{{x}}$\ 
alloys were\ computed using\ the\ GPAW code\ 
within plane\ wave basis\ set\ $\&$\ a\ PBE\ 
exchange\ correlation functional\ treatment.\
The best values\ of a lattice\ constant for\ 
the system are\ obtained\ by\ total energy\ 
calculations\ versus lattice cell\ volumes\ 
as\ fitted\ to\ the\ stabilized jellium model.\ 
Surface energies,\ cohesive energies,\ and\ 
binding energy of\ Pd$\rm_{1-x}$Ag$\rm_{{x}}$\ 
alloys were\ computed\ to\ analyze\ the\ 
stability\ properties\ of\ structures.\ 
Band structure calculations\ reveal\ the\ 
electronic\ and optical\ properties of\ 
these alloys.\ The density of states~(DOS)\ 
and projected density of\ states~(PDOS)\ show\ 
the availability\ of\ the\ eigenstates\ for\ 
occupation.\ 
The desorption\ process\ is\ studied\ within\ 
the\ Arrhenius type\ desorption rate\ $\&$\ 
a\ temperature\ programming.\ The effects of\ 
lateral interactions\ between adsorbed molecules\
on first order\ desorption (molecular adsorption)\ 
$\&$\ second order desorption\ were taken\ into\ 
account.\ Adsorption energies of H and CO on\ 
Pd$\rm_3$Ag~(111)\ as calculated using DFT\ is\ 
used\ in\ the\ process.\ The\ outcomes\ show\ 
good qualitative\ agreement with\ literature.\
\end{abstract}

\begin{keyword}
PEMFC\sep Density functional theory\
\sep Monte carlo\sep desorption rate \
\sep hydrogen\sep carbon monoxide\sep
adsorption.
\end{keyword}

\end{frontmatter}

\section{Introduction\label{sec:intro}}
Understanding the\ microscopic structure\ 
of matter,\ how daily goods are formed,\ and 
how they might be developed\ is always beneficial.\
Low-temperature\ polymer electrolyte membrane\ 
fuel cells\ (PEMFCs),\ also known as proton exchange\ 
membrane fuel cells,\ have stimulated interest in\ 
recent years as\ a feasible power source,\ 
particularly in\ automotive applications~\cite{ham2011pd}.\
PEM fuel cells\ generate electricity by combining\ 
hydrogen\ (or hydrogen-rich fuel) and oxygen\ 
(from the air),\ resulting in low pollution\ 
and high fuel economy~\cite{vielstich2009handbook}.\
The development of\ highly active and low-cost catalysts\ 
is a major challenge\ before proton exchange membrane\
fuel cells (PEMFCs)\ can find large-scale practical\ 
applications~\cite{wieckowski2003catalysis,weber2018carbon,
antolini2003formation, shao2007understanding,yu2007recent, 
chen2009shape, peng2009designer, liu2011tuning}.

Platinum has\ been widely chosen\ for a catalytic\ 
role in\ PEMFC since a\ decades ago~\cite{HL69}.\ 
However,\ there is a need to find\ replacement\ 
materials\ because Pt has\ been used as a\ 
catalyst for fuel\ cells for\ a long time\ and\ 
might face\ being\ a short\ supply in the future.\ 
Even currently, one\ of the\ major barriers\ 
for a\ cheap\ commercialization\ of\ fuel cell\ 
engines\ is the electrodes'\ high cost,\ 
which is\ mostly\ due to\ the\ use of\ 
expensive\ platinum~(Pt)-containing\ 
electro-catalysts.\ Furthermore,\ the oxygen\ 
reduction\ reaction~(ORR)\ at the cathode is\ 
the PEM fuel cell's\ rate-limiting phase,\ 
in which molecular\ oxygen is\ destroyed and\ 
mixed with protons\ and\ electrons\ supplied by\ 
the anode through\ the membrane and\ external\ 
circuit\ to form water\ as\ by-product.\ 
As a result,\ platinum-based\ catalysts can\ 
be poisoned by\ carbon monoxide~(CO)\ and other\ 
pollutants~\cite{shao2006palladium}.\ 
Because of these\ characteristics,\ pursuing for\ 
alternative\ catalyst materials\ have gotten a\ 
lot of research\ attention.\ 
A\ recent\ study\ within\ our\ research\ group~\cite{Nigussa_2019}\
has\ investigated\ palladium\ as\ a possible\ 
candidate\ for\ replacement.\ This\ corresponds\ to\ 
$\textit{\rm{x}}=0$,\ in\ the\ alloy\ composition.\ 
However,\ still\ because\ of\ the\ rare\ availability\ 
of\ Pd,\ consideration\ of\ further\ other\ 
alloys\ as\ a\ replacement\ candidate\ is\ of\ 
interest\ in\ this\ work.\ 
As a result,\ one of our primary\ motives is to\ 
investigate the properties\ of the Pd$\rm_{1-x}$Ag$\rm_{{x}}$\ 
($\textit{\rm{x}}=0.25-0.75$)\ alloy as\ components\ 
of fuel\ cell,\ as electrode.\ This\ could\ also\ 
represent\ a\ study\ on\ a\ part\ of\ a\ grand\ alloy\ 
matrix\ Pt$\rm_{1-x}$Ag$\rm_{x}$Pd$\rm_{1-x}$Ag$\rm_{x}$,\
making\ the\ fuel\ cell\ electrode,\ where\ investigating\ 
how\ certain\ region\ of\ the\ alloy\ dominated\ by\ 
Pd$\rm_{1-x}$Ag$\rm_{x}$\ behaves\ is\ the\ central\ 
interest\ point\ of\ this\ work.\ Studying\ efficiency\ 
of\ candidate\ replacement\ catalysts\ goes\ through\ 
investigating\ interactions\ of\ molecules\ 
of\ fuel\ with\ components\ of\ the\ fuel cell\  
as\ well\ as\ also\ transport\ of\ the\ molecules\ 
of\ fuel\ within\ the\ fuel\ cell.\   
 
\begin{figure}[htp!]
\centering
\begin{adjustbox}{max size ={\textwidth}{\textheight}}
\includegraphics[height=6cm,width = 9cm]{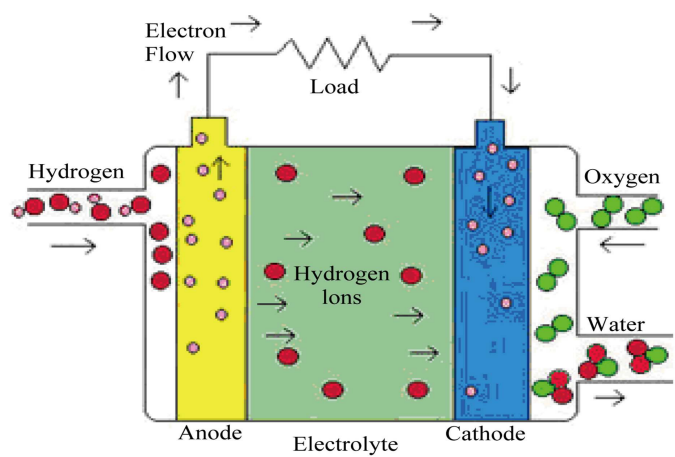}
\end{adjustbox}
\caption{Simplified diagram of the PEM fuel cell\ 
         operating process,\ taken\ 
         from~\cite{salameh2014dynamic}.\ 
         \label{fig:fig1}}
\end{figure}

The interaction of molecules\ of fuel with the fuel cell\ 
electrode\ is studied by Density\ Functional Theory~(DFT),\ 
while\ desorption of molecules\ of the\ fuel from\ the fuel\ 
cell electrode\ is studied\ by Kinetic Monte Carlo~(KMC)\ 
simulation.\ A\ Simplified diagram of the PEM fuel cell\ 
operating principles\ is shown in Fig.~\ref{fig:fig1}.\
In this paper,\ we present the results\ 
adsorption properties based\ on first\ 
principles DFT calculation,\ as well as\ 
a\ monte carlo study\ of desorption\ process\ 
from\ Pd$\rm_{1-x}$Ag$\rm_{x}$~($\rm{x}=$~0.25,\ 
0.50,\ 0.75) surface alloy where the\ DFT\ 
outputs\ are used as empirical~parameters.\
Our\ presentation\ focuses\ on\ 
Pd$\rm_{3}$Ag~(111),~specifically\
due to its stability\ compared to\ 
PdAg and PdAg$\rm_{3}$\ surfaces.\ 
The paper is organized as follows.~In the\ 
next\ section (sec.~\ref{sec:comp}),\ a\ 
detail account\ of the computational\ method\ 
is presented.\ Results\ and discussion are\ 
presented in\ section~\ref{sec:res},\ with\ 
the conclusion\ being presented in\ section~\ref{sec:conc}.
\section{Computational Methods\label{sec:comp}}
\subsection{Density Functional Theory~(DFT)}
An ab-initio simulations within gpaw code~\cite{55enkovaara2010electronic}\
is\ used\ to examine the\ electronic structure\ 
properties of the\ Palladium-Silver alloy.\
The electron wavefunction\ is approximated in gpaw\ 
using\ a projector augmented wave~(PAW) modality\ 
$\&$ expanded over\ a planewave basis set~\cite{56mortensen2005real,
57blochl1994improved}.\ The PAW technique\ is designed\ 
to successfully deal\ with the so-called pseudopotential\ 
idea,\ which is used\ in other DFT codes.\ 
The exchange-correlation\ energies are\ treated\ 
using\ PBE~\cite{24perdew1996generalized}.\
The PAW data set takes care of the electron-ion\ 
interactions.\ The\ k-points\ of\ the\ Brillouin zone~(BZ)\ are\ 
generated from the input ${\bf{k}}$-mesh\ using\ 
the\ Monkhorst-Pack scheme~\cite{59monkhorst1976special}.\
The\ number\ of valence electrons\ considered\ for\ 
each element\ within the\ paw\ data\ sets\ is\ Pd:16,\ 
Ag:17,\ H:1, C:4,\ and O:6.\ Geometry\ 
relaxations are\ carried out using BFGS\ minimizer~\cite{BS82},\ 
where optimizations\ of\ the atomic\ coordinates\ 
and the unit\ cell degrees of freedom\ is done\ 
within\ the concept\ of the Hellmann-Feynman forces\ 
and\ stresses~\cite{PRF39, NM85}\ as\ calculated on\ 
the Born-Oppenheimer~(BO) surface~\cite{WM91}.\ 
The convergence\ criteria for the forces\ were set\ 
at 0.05 eV/{\AA}.

Surfaces\ are\ modelled\ by\ a\ slab\ containing\ 
four\ atomic layers of\ the\ desired\ $hkl$\ indices.\ 
The\ top two atomic\ layers are\ allowed\ to relax\ 
upon\ geometry\ optimizations\ while\ the\ bottom\ 
two\ layers\ are\ frozen\ to\ a\ bulk\ geometry.\ 
A\ vacuum\ layer\ of\ 10~{\AA}\ is\ added\ to\ 
the\ slabs\ making\ a\ supercell\ which\ avoids\ 
lateral\ interaction\ between\ slabs,\ while\ 
also\ mimicks\ a\ three\ dimensional cell.\ 
Free\ adsorbate\ molecules/atoms\ prior\ to 
adsorption are\ relaxed\ in\ a\ cubic\ unit cell\ 
of\ side length\ 10~{\AA}.\ Adsorptions\ are\ done\ 
by\ putting\ adsorbate\ molecules / atoms\ on\ the\ 
top\ most\ atomic\ layers\ of\ the\ slabs\ $\&$\ 
then\ allowing\ geometry\ relaxations\ to\ take\ 
place.\ Adsorption\ energies\ are\ calculated\ 
according\ to:\ 
\begin{equation}
 E_{ads} = -(E_{SM} - E_S - E_M) \label{eads}
\end{equation}
Here,\ $E_S$ denotes total\ energy of a clean surface,\ 
$E_M$\ denotes total energy\ of an isolated molecule,\ 
and $E_{SM}$ denotes\ total energy of a relaxed geometry\ 
containing both\ substrate and molecule.\
All energies are per supercell.\ Accordingly to\ 
Eq.~\eqref{eads},\ thus,\ positive values\
in energies indicate\ exothermic reactions\ 
while negative\ values denote endothermic.\
A\ cubic\ unit cell\ with 3~Pd and 1~Ag\ atoms\ 
has been\ considered for Pd$\rm_{3}$Ag bulk\ 
structure.~For\ PdAg structure, the\ cubic\ unit\ 
cell with 2~Pd and 2~Ag\ atoms\ has been\ used.\ 
For\ PdAg$\rm_{3}$\ structure,\ the cubic unit\ 
cell with 1~Pd and 3~Ag\ atoms\ has\ been\ used.\ 
Since\ plane wave basis set\ is used\ for the\ 
expansion\ of electron\ wavefunctions,\ the\ 
k-points $\&$\ cut-off\ energies\ chosen\ would\ 
impact\ on\ total\ energy\ of\ the\ systems.\ 
Our\ calculations\ show\ that\ 
{\bf{k}}-mesh of 8{$\times$}8{$\times$}8\ can\
be\ optimum\ in\ the\ bulk\ calculations.\
With a fixed {\bf{k}}-mesh of $8{\times}8{\times}8$,\
a\ convergence test\ of the total energy\ with\ 
respect to\ energy cutoff\ is performed.\ 
As a result,\ we have chosen\ a\ cut-off\ 
energy~(ecut)~of\ 600 eV\ as\ an\ optimum\ 
in\ the\ rest\ of\ our\ calculations.\ With\ 
surface\ calculations,\ correspondingly,\ 
a\ {\bf{k}}-mesh of\ 8{$\times$}8{$\times$}1\ 
and an ecut of 600~eV\ is\ used.\ 

\subsection{Kinetic Monte Carlo~(kMC)}
In\ experiments,\ thermodynamic\ and kinetic\ 
parameters\ of\ desorption\ processes or\ 
decomposition\ reactions\ can be\ determined\ 
using\ temperature programmed desorption~(TPD).\ 
In simulations,\ there are\ 
numerous approaches\ for simulating\ TPD\ 
spectra,\ with Monte Carlo\ simulation\ 
being one of\ the most useful\ and crucial.\
This approach\ has the distinct\ advantage\ 
of requiring\ as fewest assumptions as\ 
possible.\ The probability\ for a particle at\ 
site $i$ to\ desorb in\ the interval~between\ 
$t$,~$\&$\ ($t$ + $\Delta$t)~may be\ calculated\ 
directly\ using Eq.~\eqref{rd1},\ 

\begin{equation}
 r\rm_{d} = \nu \theta^n exp\left(-\frac{E\rm_{\text{ads}}}{RT}\right)
\label{rd1} 
\end{equation}
Where $\nu = 10^{13} s^{-1}$,\ is the vibrational\ 
pre-factor\ or pre-exponential\ factor.\ R and T\ 
are\ universal\ gas constant and absolute\ temperature,\ 
respectively.\ The\ temperature\ in the simulation\ 
is updated\ by \ Eq.~\eqref{beta}.\ 
\begin{equation}
T = T\rm_{0} + {\beta}\hspace{0.2mm}t
\label{beta}
\end{equation}
where\ $t$\ is\ time for a\ desorption event and 
$\beta$ is the heating rate.\
The probability\ for a particle at\ site $i$ to\ 
desorb in\ the interval~between\ $t$,~$\&$\ 
($t$ + $\Delta$t) when lateral\ interaction is\ 
considered\ is given by Eq.~\eqref{rd},\
\begin{equation}
 r_d = -\frac{d\theta}{dt} = \nu \theta^n 
 exp\left(-\frac{E\rm_{\text{ads}}
 +E\rm_{\text{int}}(\theta)}{RT}\right) 
 \label{rd}
\end{equation}
$\nu$ is\ the so-called\ pre-exponential factor,\ 
$n$ is\ the reaction order\ or kinetic desorption\ 
order~(with values~0, 1, and 2),\ $E\rm_{\text{ads}}$\ 
is\ the activation energy\ for desorption (for this\ 
work taken from DFT),\ $E\rm_{\text{int}}$ is\ 
the interaction\ energies between\ the adsorbate\ 
molecules,\ and $R$ and T\ are\ the universal\ 
gas\ constant and sample\ surface\ temperature,\ 
respectively.\ The\ interaction\ term\ 
of\ Eq.~\eqref{rd}\ is\ approximated\ according\ 
to
\begin{equation}
E\rm_{int}={\sum\limits_{k}}{N\rm_{kn}}E\rm_{kn}
\label{eqint2}
\end{equation}
with\ $E\rm_{kn}$~(pair-interaction\ energy)\ 
values\ are\ taken\ from\ literature~\cite{sraaen}.\ 
$N\rm_{kn}$~denotes\ number\ of\ $k\rm^{th}$\ 
nearest\ neighbor.\ For\ the\ Markovian\ 
process,\ the\ metropolis\ algorithm\ within\ 
the\ concept\ of\ Meng~and~Weinberg~\cite{BW94}\
is\ applied.\

\section{Results and discussion \label{sec:res}}
\subsection{Equation of state~(EOS) and cohesive energy}
We determine\ the equilibrium lattice\ constant\ 
and\ bulk modulus~(material's\ resistance to\ 
external\ pressure),\ using the equation\ of\ 
state curve-fitting\ approach\ for\ different\ 
palladium-silver\ bulk\ structures.\ The\ EOS\ 
is\ obtained\ by\ a\ curve\ fits\ within\
Stabilized Jellium model~(sjeos)~\cite{APBAF2003}\
to\ an\ energy versus volume\ calculation\ data,\ 
which\ also\ additionally\ outputs\ 
other\ important\ bulk structural\ quantities,\ 
such\ as minimum volume\ and\ minimum energy.\ 
This fit approach\ seems to yield a\ bulk modulus
values which\ is more\ closer to\ experiments\ 
than\ with using\ other fit methods\ 
such as Birch-Murnaghan~\cite{FDM44,FB78}~approach.\
However,\ the predicted values of the optimum\ 
lattice\ constants by using all\ the fit methods\ 
have\ appeared to\ be identical.\ 
The\ output\ parameters\ obtained\ within\ the\ 
sjeos\ fit\ approach,\ along\ with\ comparison\ 
with\ experiment\ results,\ is\ 
presented\ in\ Table~\ref{tab1}.\ The experimental\ 
results are\ taken from\ literatures~\cite{lovvik2002density, 
coles1956lattice, kittel1996introduction,chan1969thermodynamic}.\

The cohesive\ energy~[eV/atom]~is\ defined\ as\ the\ 
energy\ required\ to\ dissociate\ the\ alloy\ 
compound into~free~(neutral)\ atoms.\ It\ is\
calculated\ according\ to\
\begin{equation}
E\rm_{coh} = {\sum\limits_{\rm_{atom}}}{E\rm_{atom}}-
\frac{1}{N}{E\rm_{bulk}}
\end{equation}
where $E\rm_{coh}$ is the\ cohesive energy,\ 
$N$\ is\ the number\ of atoms\ in\ the\ unit\ 
cell\ of\ the\ palladium-silver\ alloy.\ 
$E\rm_{atom}$\ is the atomic\ energy of a\ 
free\ palladium and silver\ atoms.\  
Accordingly,\ as\ also\ shown\ in\ Table~\ref{tab1},\
the\ cohesive energy of Pd$\rm_{3}$Ag $>$ PdAg $>$ PdAg$\rm_{3}$.\
That\ means\ Pd$\rm_{3}$Ag\ is\ more\ stable.\\ 
\begin{table*}[htp!]
\addtolength{\tabcolsep}{4.8mm}
\renewcommand{\arraystretch}{2.2}
\centering
\caption{The lattice constant $a$~[{\AA}],\ the bulk\ 
         modulus B~[GPa],\ formation energy~$E\rm_{f}$~[eV/atom],\ 
         and\ the cohesive energies~$E\rm_{coh}$~[eV/atom]~of\ 
         selected\ metals\ $\&$\ alloy\ systems\ of\ this\ 
         work.\label{tab1}}
   
\begin{tabular}{lcccccc}
\hline
\multirow{2}*{Quantity} & \multirow{2}*{Source} & \multicolumn{5}{c}{System}\\
\cline{3-7}
{} & {} & Pd & Pd$\rm_{3}$Ag & PdAg & PdAg$\rm_{3}$ & Ag\\
\hline
\multirow{2}*{$a$} & {This work} & 3.94 & 3.98 & 4.02 & 4.07 & 4.07\\
                 & {Exp.} & 3.88 & 3.92 & 3.97 & 4.02 & 4.07\\
\hline
\multirow{2}*{B} & {This work} & 188 & 139.5 & 127 & 103.7 & 109\\
                  & {Exp.} & 180.8 & 139.4 & 116 & 103.1 & 100\\ 
\hline
\multirow{2}*{$E\rm_{f}$} & {This work} & - & 3.08 & 3.31 & 3.61 & -\\  
                        & {Exp.} & - & - & - & - & -\\
\hline
\multirow{2}*{$E\rm_{coh}$} & {This work} & - & 3.31 & 2.89 & 2.60 & -\\  
                        & {Exp.} & - & 3.60 & 3.37 & 3.16 & -\\                           
\hline
\end{tabular}
\end{table*}
The energy\ necessary to\ form the\ 
alloy compound\ from\ its separate\ elemental 
bulk structures\ is represented by\ 
formation\ energy,~$E\rm_{f}$~[eV/atom],~which\ 
is\ given\ by\ 
\begin{equation}
E\rm_{f}=\frac{1}{N}{E\rm_{bulk}} 
- {\sum\limits_{i}}\frac{1}{N\rm_{i}}{E\rm_{b}^{i}}
\end{equation}
where\ $N$~is\ number\ of\ atoms\ in\ the\ 
unit\ cell\ of\ the\ alloy\ compound,\ 
$E\rm_{bulk}$\ is\ total\ energy\ of\ 
the\ bulk\ alloy\ compound,\ $E\rm_{b}^{i}$\ is\ 
total energy\ of\ bulk\ of\ element\ $i$,\ 
$\&$~$N\rm_{i}$\ is\ number\ of\ atoms\ 
in\ the\ unit\ cell\ of\ the\ elemental\ 
bulk\ forms.\ In terms of\ formation energy,\ 
thus,\ $E\rm_{f}$\ of\ Pd$\rm_{3}$Ag $<$ PdAg $<$ PdAg$\rm_{3}$\
(see Table~\ref{tab1}).\ This\ means\ Pd$\rm_{3}$Ag\
alloy\ is\ relatively\ a\ more\ favorable\ 
to\ be\ formed.\ 
From\ the\ EOS\ output,\ it is~investigated\ 
that\ the total energy per atom\ values\ 
are\ about\ -3.6~eV,~-3.4~eV,~-3.1~eV,\ respectively,\ 
for Pd$\rm_3$Ag,~PdAg,~$\&$~PdAg$\rm_{3}$~structures.\ 
Furthermore,~the\ bulk modulus~(B)~value\ for Pd$\rm_{3}$Ag\
$>$ PdAg $>$ PdAg$\rm_{3}$\ means\ that\ Pd$\rm_{3}$Ag\
has\ relatively\ better\ operating~$\&$~endurance\
capacity\ under\ extreme\ pressures.\ The\ fact\ 
that\ the\ optimal\ geometry\ structure\ properties\ 
presented\ in\ Table~\ref{tab1}~compare\ reasonably\ 
well\ with\ experiments\ also\ gives\ the\ authors\ 
a\ full\ confidence\ on\ the\ validity\ of\ the\ 
chosen\ model\ alloy\ compound~$\&$~surfaces\ of\ 
this\ study.\ 

\subsection{Surface energy}
\begin{figure*}[ht!]
\centering
\begin{adjustbox}{max size ={\textwidth}{\textheight}}
(\centering a){{\includegraphics[scale=0.1]{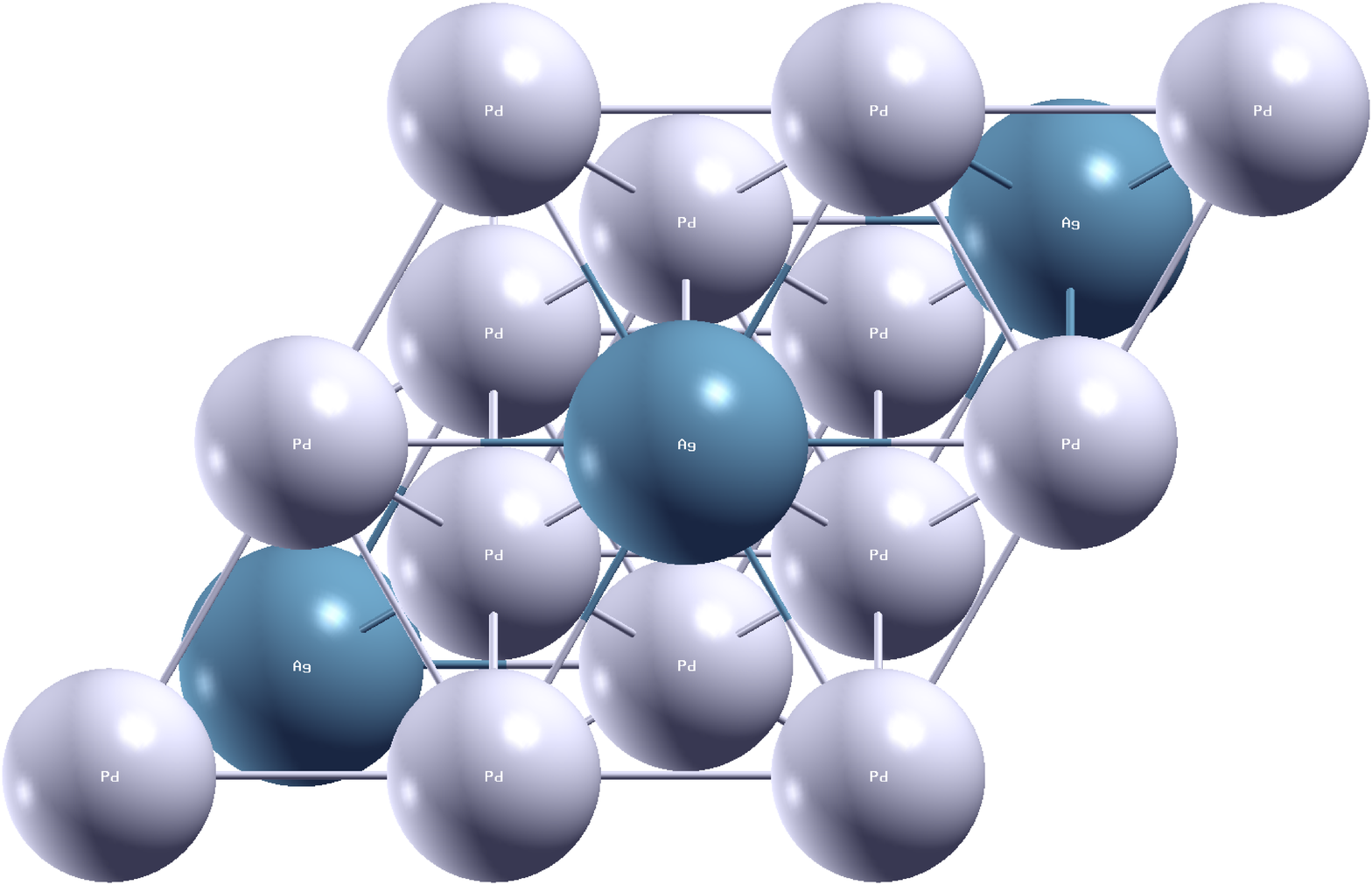}}}
\qquad
(\centering b) {{\includegraphics[scale=0.12]{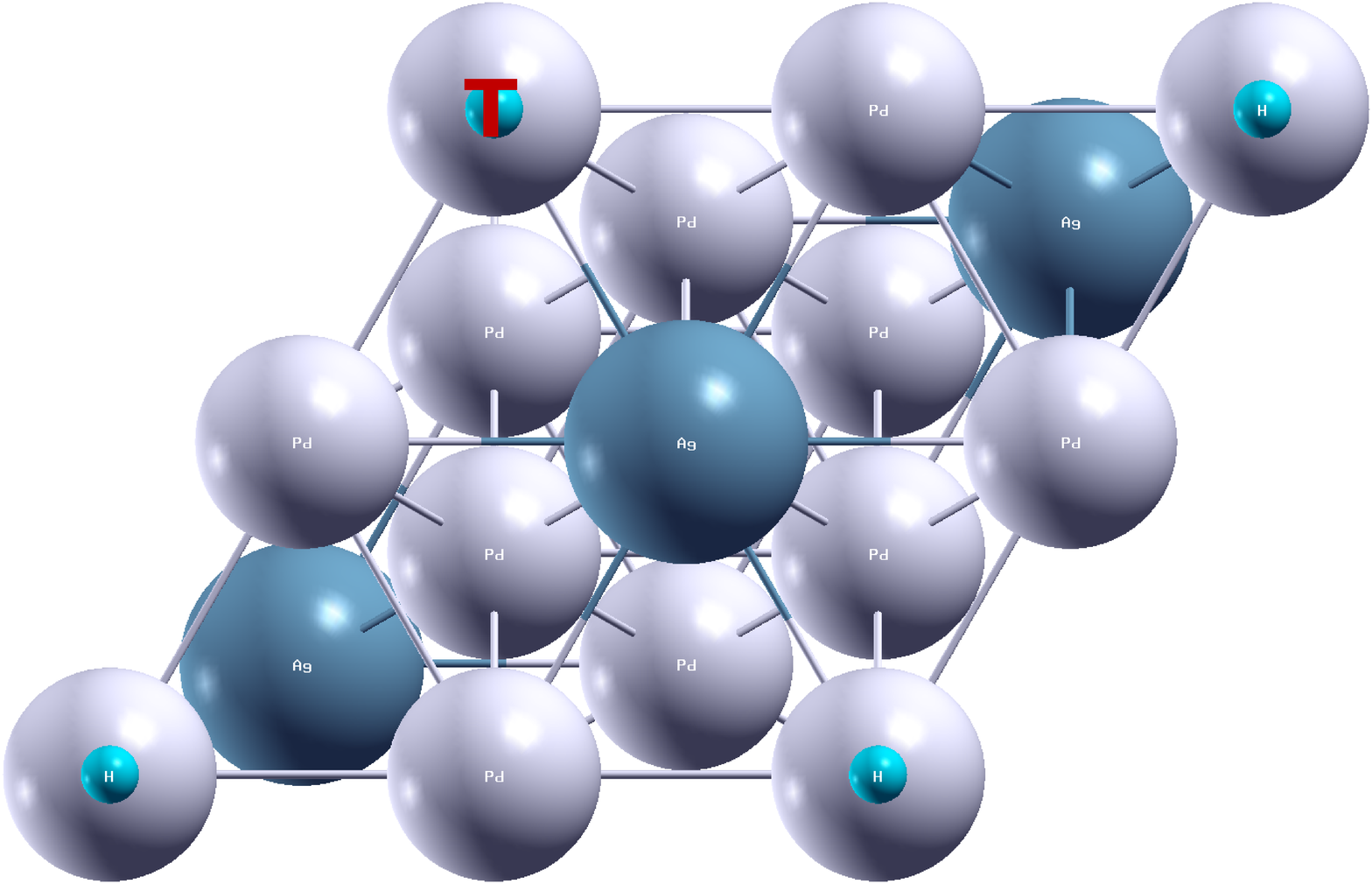}}}
\qquad
(\centering c){{\includegraphics[width=6cm]{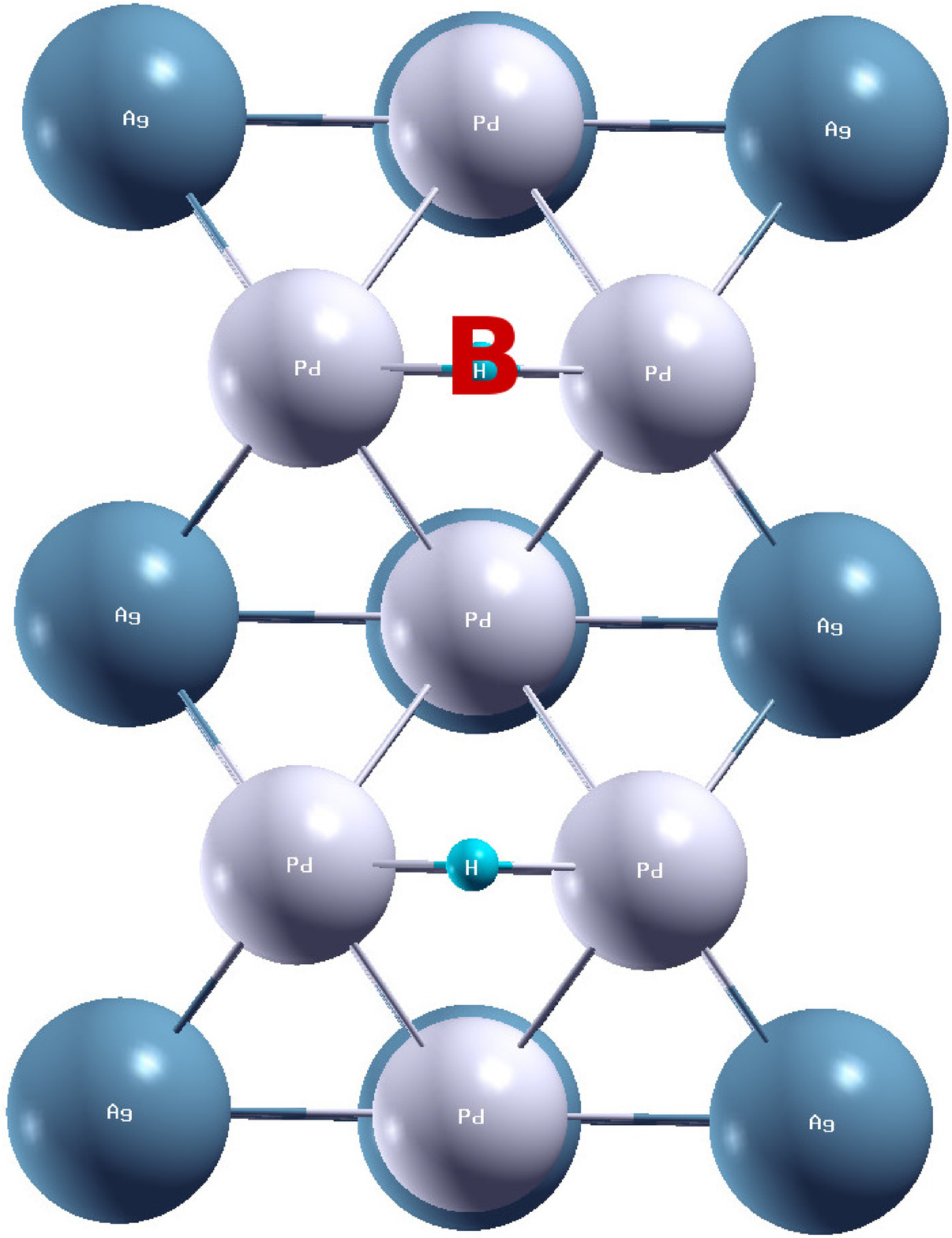}}}
\qquad
(\centering d) {{\includegraphics[width=6cm]{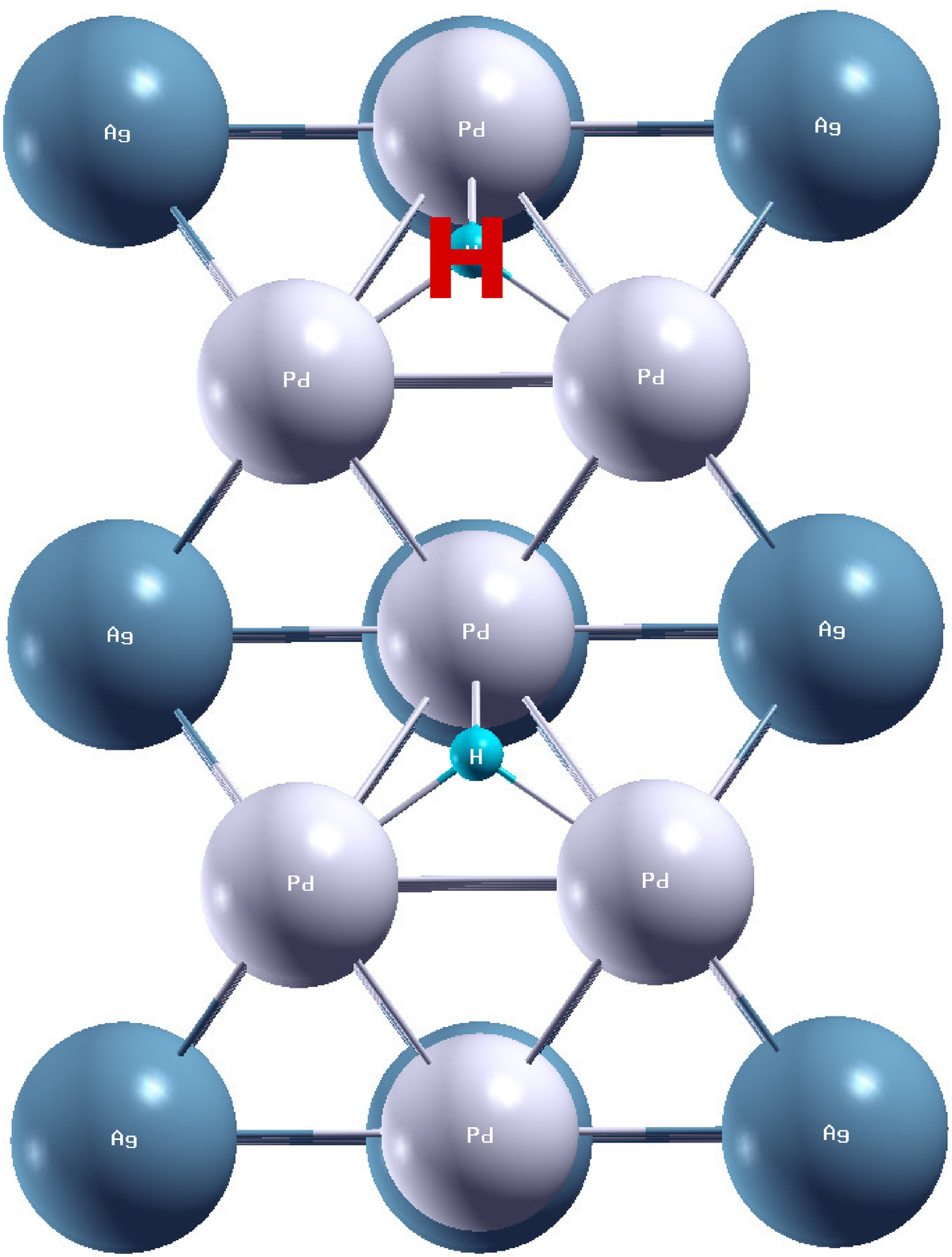}}}
\end{adjustbox}
\caption{(a) Pd$\rm_{3}$Ag~(111) surface and\ 
         (b) Pd$\rm_{3}$Ag~(111)~top~(T)~site view.
         (c) Pd$\rm_{3}$Ag~(111)~bridge~(B)~site view\ and\ 
         (d) Pd$\rm_{3}$Ag~(111)~hollow~(H)~site view.
         Color online. Colors:~Pd-light~grey,~H-cyan,\
         Ag-dark~cyan.\label{fig2}}

\end{figure*}

Surface energy is used to characterize\ 
the surface of a substrate.\ Its values\ range\ 
from high to low.\ Adhesion between\ unlike materials\ 
is determined by\ the molecular force\ of attraction\ 
between them.\ The strength of\ the attraction is\ 
determined by\ the substrate's surface energy.\
A strong molecule\ attraction is\ indicated by\ 
a high surface energy,\ whereas a weak molecular\ 
attraction is\ indicated by\ a low surface energy.\
A high surface energy\ would\ also indicate\ 
increased\ reactivity\ with\ adsorbates,\ while\ 
low\ surface\ energy\ would\ indicate\ increased\ 
stability.\ Surface characteristics\ are\ extremely\ 
important,\ especially in catalysis.\ Generally,\ 
the surface\ energy~($\sigma$~[eV/{\AA}$\rm^{2}$])\ 
is defined\ as\
\begin{equation}
\sigma = \frac{1}{2A}\left(E\rm_{slab} -
\frac{N\rm_{slab}}{{N\rm_{bulk}}} E\rm_{bulk}\right)
\end{equation}
where~$E\rm_{slab}$ denotes the total\ energy of a slab\ 
unit cell,\ $N\rm_{slab}$ means number\ of atoms in the\ 
slab unit cell,\ $N\rm_{bulk}$ means number\ of atoms\ 
in the bulk\ unit\ cell,\ and $E\rm_{bulk}$\ is the\ 
total energy of\ a bulk unit~cell.\ $A$ is the\ surface\ 
area of\ a slab\ unit cell,\ and the\ factor\ half\ 
accounts\ for the top\ and bottom surfaces.\\
\begin{table*}[htp!]
\addtolength{\tabcolsep}{12mm}
\renewcommand{\arraystretch}{2.2}
\centering
\caption{Surface energies of Pd$\rm_{1-x}$Ag$\rm_{{x}}$~surfaces,\
         $\text{x}$~goes from 0.25~to~0.75.~Values~corresponding\
         to\ 100,~110,~$\&$~111 facets\ are\ given.\label{tab2}}

\begin{tabular}{lccc}
\hline
\multirow{2}*{System}{} & \multicolumn{3}{c}{{Surface energy~($\sigma$)~[eV/{\AA}$\rm^{2}$]}}\\ 
\cline{2-4}
{} & {100} & {110} & {111}\\ 
\hline
Pd$\rm_3$Ag & 0.085 & 0.089 & 0.072\\  
PdAg & 0.076 & 0.078  & 0.062\\ 
PdAg$\rm_3$ & 0.056  & 0.060 & 0.049\\  
\hline
\end{tabular}
\end{table*}
As\ shown\ in\ Table~\ref{tab2},\ the surface\ 
energies\ for the surface facets increase according\ 
to\ $(111)<(100)<(110)$\ for\ all\ the\ structures.\ 
This\ means\ (111)~surfaces\ offer\ a\ relatively\ 
more\ stable\ geometries,\ while\ (110)\ facets\ 
would\ likely\ be\ expected\ to\ be\ more\ reactive\ 
to\ impurities / adsorbates.\ However,\ the\ 
surface\ energies\ values,\ when\ compared\ 
across\ the\ different\ structures,\ are\ closely\ 
related,\ albeit\ the\ small\ differences\ can\ be\ 
taken\ as\ being\ a\ marginal\ errors.\ Studies of\ 
interactions of\ these surfaces\ with\ 
adsorbates:~hydrogen~(H),\ and carbon monoxide~(CO)\ 
are discussed in\ the next~subsections.\ 
The adsorption site\ terminologies of\ top~(T),\ 
bridge~(B),\ and hollow~(H)~refer to the\ 
placement\ of adsorbates\ at the geometries\ 
depicted in Fig.~\ref{fig2}.\
The corresponding\ adsorption results are\ 
shown in\ Tables~\ref{tab3}~$\&$~\ref{tab4}.\

\subsection{Adsorption~energy~of~Hydrogen~(H)~and~Carbon~monoxide~(CO)} 
Upon\ the\ adsorptions,~a CO molecule\ or H is placed\ 
on\ the surfaces on\ one side of\ the slab at\ either the\ 
on-top~(T) site,\ on bridge~(B),\ or fcc hollow~(H)~site.\
The adsorption energies\ are calculated as\
\begin{eqnarray}
 E\rm_{ads} &=& -\left(E\rm_{slab + CO} - E\rm_{slab} 
 - E\rm_{CO}\right)  \hspace{4mm}   \text{for~CO}\\
 E\rm_{ads}& = & -\left(E\rm_{slab + H} - E\rm_{slab} 
 - 0.5\times E\rm_{H\rm_2}\right)  \hspace{4mm}  \text{for~H}
\end{eqnarray}
where $E\rm_{CO}$ is\ the energy of an isolated\ 
(free)~$\rm{CO}$ molecule,\ and $E\rm_{H\rm_{2}}$ is\ 
the energy of\ an isolated~H$\rm_{2}$\ molecule.\ 
The results are\ given in Tables~\ref{tab3}~$\&$~\ref{tab4}.\ 
\begin{table*}[ht!]
\addtolength{\tabcolsep}{9.0mm}
\renewcommand{\arraystretch}{1.3}
\centering
\caption{Adsorption energies\ of hydrogen~(H)\ on different\ 
         surface faces~(i.e.,\ 100,\ 110,\ $\&$\ 111)~and change\ 
         in charge of\ adsorbate for selected\ 
         system~(-~sign indicates a loss) to palladium-silver\ 
         alloy~system~(Pd$\rm_{1-x}$Ag$\rm_{x}$,\ 
         $\text{x}$~goes~from~0.25~to~0.75).\
         The\ adsorptions\ correspond\ to\ adsorbate\ 
         coverage of 1 adsorbate species per slab\ 
         unit cell~(1~ML).\label{tab3}}

\begin{tabular}{lcccc}
\hline
{System} & {Surface} & {Site} & E$\rm_{ads}$~[eV] & ${\Delta}Q\rm_{H}~[e]$\\ 
\hline
\multirow{9}*{Pd$\rm_{3}$Ag} & \multirow{3}*{100} & T & -0.81 & {-}\\
           {} & {} & B & 0.15 & -0.06\\
           {} & {} & H & 0.32 & {-}\\ 
           \cline{3-5}
           {} & \multirow{3}*{110} & T & -0.34 & {-}\\
           {} & {} & B & -0.15 & -0.06\\
           {} & {} & H & -0.52 & {-}\\
           \cline{3-5}
           {} & \multirow{3}*{111} & T & -0.34 & {-}\\
           {} & {} & B & 0.34 & -0.16\\
           {} & {} & H & 0.59 & -0.10\\
           \cline{2-5}
                            
\multirow{9}*{PdAg} & \multirow{3}*{100} & T & -0.84 & {-}\\           
                    {} & {} & B & -0.34 & -0.11\\
                    {} & {} & H & -0.29 & {-}\\
                    \cline{3-5}
                    {} & \multirow{3}*{110} & T & -0.40 & {-}\\
                    {} & {} & B & 0.14 & -0.14\\
                    {} & {} & H & -0.27 & {}\\
                    \cline{3-5}
                    {} & \multirow{3}*{111} & T & -0.40 & -0.09\\
                    {} & {} & B & 0.02 & -0.14\\
                    {} & {} & H &  0.34 & -0.06\\
                    \cline{2-5}
                    
\multirow{9}*{PdAg$\rm_{3}$} & \multirow{3}*{100} & T & -0.88 & {-}\\
                             {} & {} & B & -0.39 & -0.09\\
                             {} & {} & H & -0.57 & {}\\
                             \cline{3-5}
                             {} & \multirow{3}*{110} & T & -0.97 & {-}\\
                             {} & {} & B & -0.37 & {-}\\
                             {} & {} & H & -0.31 & {-}\\
                             \cline{3-5}
                             {} & \multirow{3}*{111} & T & -0.34 & -0.09\\
                             {} & {} & B & 0.04 & -0.16\\
                             {} & {} & H & 0.08 & -0.19\\
\hline
\end{tabular}
\end{table*}
\begin{table*}[ht!]
\addtolength{\tabcolsep}{9.0mm}
\renewcommand{\arraystretch}{1.3}
\centering
\caption{Adsorption energies\ of carbon monoxide~(CO)\ on different\ 
         surface faces~(i.e.,\ 100,\ 110,\ $\&$\ 111)~and change\ 
         in charge of\ adsorbate for selected\ 
         system~(+~sign indicates a gain) from palladium-silver\ 
         alloy~system~(Pd$\rm_{1-x}$Ag$\rm_{x}$,\ 
         $\text{x}$~goes~from~0.25~to~0.75).\
         The\ adsorptions\ correspond\ to\ adsorbate\ 
         coverage of 1 adsorbate species per slab\ 
         unit cell~(1~ML).\label{tab4}}

\begin{tabular}{lcccc}
\hline
{System} & {Surface} & {Site} & E$\rm_{ads}$~[eV] & ${\Delta}Q\rm_{CO}~[e]$\\ 
\hline

\multirow{9}*{Pd$\rm_{3}$Ag} & \multirow{3}*{100} & T & 0.09 & {}\\
           {} & {} & B & 0.67 & +0.16\\
           {} & {} & H & -1.62 & {-}\\ 
           \cline{3-5}       
           {} & \multirow{3}*{110} & T & 1.66 & {-}\\
           {} & {} & B & 1.68  & +0.15\\
           {} & {} & H & -0.95 & {-}\\
           \cline{3-5}
           {} & \multirow{3}*{111} & T & 1.35 & +0.16\\
           {} & {} & B & 1.26  & +0.15\\
           {} & {} & H & 0.83 & {}\\
           \cline{2-5}
                           
\multirow{9}*{PdAg} & \multirow{3}*{100} & T & 0.06 & {-}\\           
                    {} & {} & B & -0.78 & +0.26\\
                    {} & {} & H & -2.89 & {-}\\
                    \cline{3-5}
                    {} & \multirow{3}*{110} & T & 1.54 & {-}\\
                    {} & {} & B &  0.55 & +0.18\\
                    {} & {} & H & -2.18  & {}\\
                    \cline{3-5}
                    {} & \multirow{3}*{111} & T & 1.37 & +0.18\\
                    {} & {} & B &  0.15 & +0.25\\
                    {} & {} & H & -0.01 & {-}\\
                    \cline{2-5}
                  
\multirow{9}*{PdAg$\rm_{3}$} & \multirow{3}*{100} & T & -0.20 & {-}\\
                             {} & {} & B & -0.80 & +0.15\\
                             {} & {} & H & -3.06 & {}\\
                             \cline{3-5}
                             {} & \multirow{3}*{110} & T & -0.12 & {-}\\
                             {} & {} & B & -1.07 & {-}\\
                             {} & {} & H & -4.25 & {-}\\
                             \cline{3-5}
                             {} & \multirow{3}*{111} & T & 1.33 & +0.19\\
                             {} & {} & B & 0.10 & +0.30\\
                             {} & {} & H & -0.92 & {-}\\
\hline
\end{tabular}
\end{table*}
The\ results\ in\ the\ tables\ show\ that\ 
Pd$\rm_{3}$Ag~(111)~shows\ modest\ reactivity\ 
towards\ both\ adsorbates,~while~Pd$\rm_{3}$Ag~(110)\ 
looks\ to\ be\ more\ reactive\ with\ CO~(albeit\ 
it\ may\ result\ in a\ poisoning\ effect).\
However,\ Pd$\rm_{3}$Ag~(111)\ is\ also\ 
a\ more\ favored\ surface for consideration,\ 
since\ it\ shows\ a\ relatively\ more\ stability.\ 
The\ charge\ differences\ ${\Delta}Q$\ are\
computed\ according\ to\ 
\begin{equation}
{\Delta}Q = e{\int}{\Delta}n(r){d\rm^{3}}r
\label{eqdelq}
\end{equation}
where\ ${\Delta}n(r)$\ is\ electron\ density\ 
differences obtained\ from the\ ground state\ 
calculations.\ 
The\ charge\ differences\ are\ calculated\ according\ 
to\ the\ concept\ in\ Eq.~\eqref{eqdelq}~$\&$~using\ 
Bader\ charge\ formulation~\cite{RB90}\ 
as implemented\ by\ literature~\cite{TSH2009}. 
Hydrogen\ seems\ to\ release\ charges\ 
to\ the\ surfaces,\ while\ CO\ gains.\ 
These values~(see~Table~\ref{tab3})\ show\ a\ 
donation\ of\ 0.16$e$~per\ one\ hydrogen\ atom\ 
to\ the\ system.~CO\ gains\ nearly\ about\ the\ 
same\ from\ the\ system~(see~Table~\ref{tab4}).\  

\subsubsection{Adsorption\ energies\ on\ Pd$\rm_{3}$Ag~surfaces}
On~(100)~surface,\ an adsorption energy\ of up to\ 
0.32~eV~is\ investigated\ upon\ a\ dissociated\ 
adsorption\ as hydrogen atoms\ at the H site.\ At\ 
such\ a\ higher\ adsorbate\ coverages\ as 1~ML,\ 
the\ energetically\ favorable adsorption\ sites\ 
are H~$\&$~B,\ with\ a\ relative\ favorability\ 
order\ being~H~$>$~B.\ In\ the\ case\ of\ CO\
adsorption,\ the\ adsorption\ energy is\ 
up\ to\ 0.67~eV,~$\&$~the\ favored\ sites\ 
are\ T~$\&$~B,~with\ the\ order\ being B~$>$~T.\\
On~(110)~surface,~adsorption\ of\ hydrogen\ atoms\ 
seems\ to\ be\ endothermic\ at\ all\ sites,\ 
while\ adsorption\ of\ CO\ can have\ adsorption\ 
energy\ of\ up to\ 1.68~eV,\ with\ favorable\
sites\ being\ T~$\&$~B,\ and\ the\ order\
of\ favorability\ being\ B~$>$~T.\ 
On~(111)~surface,\ a dissociated adsorption\ 
as\ hydrogen atoms\ is\ preferred\ with\ 
adsorption\ energies\ of\ up to 0.59~eV,\ 
as shown in the\ Table~\ref{tab3}.\ The\ 
favorable\ sites\ are\ H $\&$ B,\ with\ the\ 
order\ of\ relative\ favorability\ being\ 
H~$>$~B.\ With\ adsorption\ of\ CO,\ all\
sites\ are\ favorable\ with\ the\ adsorption\ 
energy of up to 1.35~eV,~$\&$~the\ relative\ 
favorability\ order\ of\ the\ sites\ 
seems\ as~T~$>$~B~$>$~H.\ The\ desorption\ energy\ 
values on\ Pd$\rm_{3}$Ag~(111)\ is\ comparable\ 
to\ but\ slightly\ smaller\ than\ a\ counterpart\ 
on\ Pt~(111)~\cite{KN2020},~means\ an\ improved\
performance\ of\ the\ model\ for\ the\ 
desired\ application.~Smaller\ adsorption\ 
energy\ of hydrogen\ would\ be\ good\ to\ 
a\ fuel\ cell\ operation\ at\ lower\ 
temperatures,\ while\ smaller\ adsorption\ 
energy\ of\ CO means\ good\ chance\ for\
reducing\ CO\ poisoning.\    
\subsubsection{Adsorption\ energies\ on\ PdAg~surfaces}
With\ adsorption\ of\ hydrogen\ at\ a\ coverage\ 
of\ 1~ML,\ the\ B\ site\ of\ (110)~and\
the\ B~$\&$~H~sites~of\ (111)\ surface\ 
are\ favorable\ adsorption\ sites.~The\ adsorption\
energies\ of\ up to~0.34~eV,\ which occurs\ at\ 
H~site\ of\ (111)\ is\ investigated~(Table~\ref{tab3}).\\
With\ adsorption\ of\ CO,\ T~site\ of\ (100),\
T~$\&$~B\ sites\ of\ (110),\ and T~$\&$~B\ sites\ 
of\ (111)\ are\ the\ favored\ adsorption\ 
sites,\ with\ adsorption\ energies\ of\ up to\
1.54~eV,\ which\ occurs\ at\ T\ site\ of\ 
(110)~surface.\ 
\subsubsection{Adsorption\ energies\ on\ PdAg$\rm_{3}$~surfaces}
At\ such\ a\ higher\ adsorbate\ coverage\ of\ 
1~ML,~hydrogen\ adsorbs\ on\ (111)\ surface\ 
at\ a\ favorable\ sites\ of\ B~$\&$~H,\ 
with\ adsorption\ energy\ of\ up to\ 0.1~eV.\ 
Adsorption\ of\ CO\ happens\ at\ a\ favorable\ 
sites\ of\ T~$\&$~B,\ with\ adsorption\ 
energies\ of\ up to\ 1.3~eV~at\ the\ T~site.\  

\subsection{Electronic Properties} 
\subsubsection{Density of States~(DOS) and Band structures}
\begin{figure*}[htp!]
\centering
\begin{adjustbox}{max size ={\textwidth}{\textheight}}
(\centering a)~{{\includegraphics[scale=0.9]{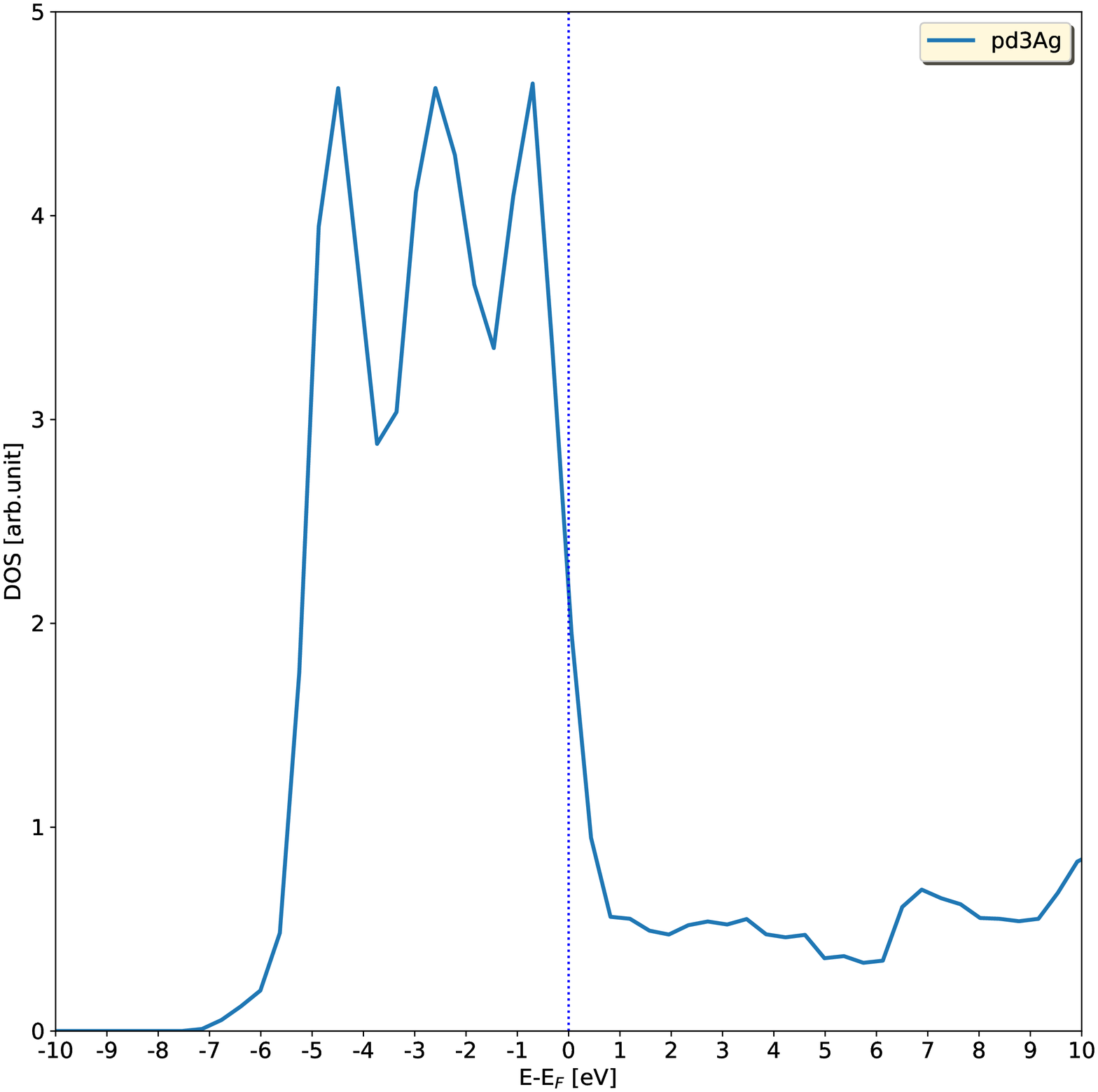}}}
\qquad
(\centering b)~{{\includegraphics[scale=0.9]{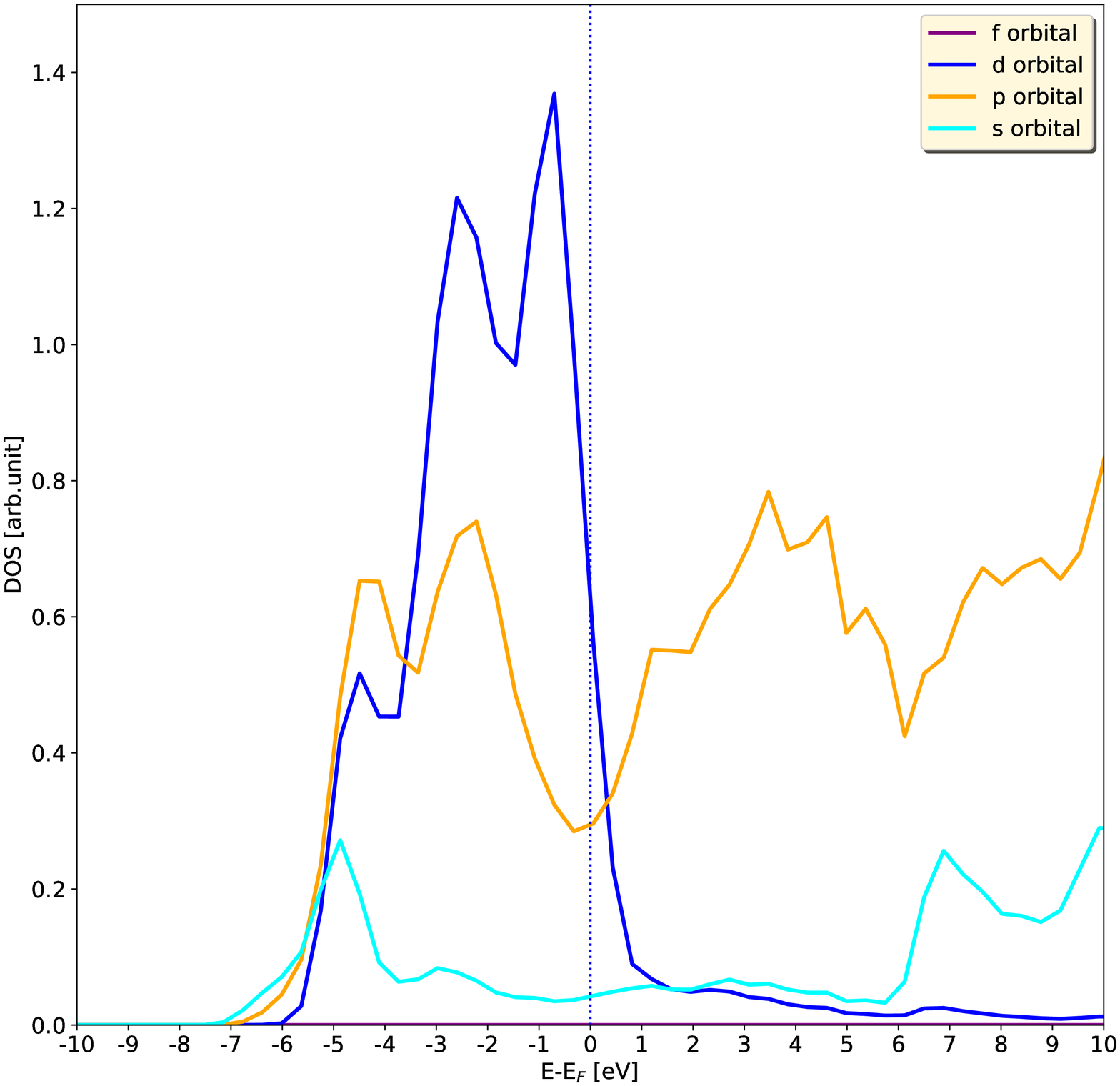}}}
\end{adjustbox}
\caption{(a)~Density of states of bulk~Pd$\rm_{3}$Ag.~(b)~Projected
 density of states of bulk~Pd$\rm_{3}$Ag\label{fig34}}
\end{figure*}

The\ DOS~$\&$~PDOS\ curves 
are\ presented\ in~Fig.~\ref{fig34}.\
\begin{figure*}[htbp!]
\centering
\begin{adjustbox}{max size ={\textwidth}{\textheight}}
\includegraphics[scale=1.0]{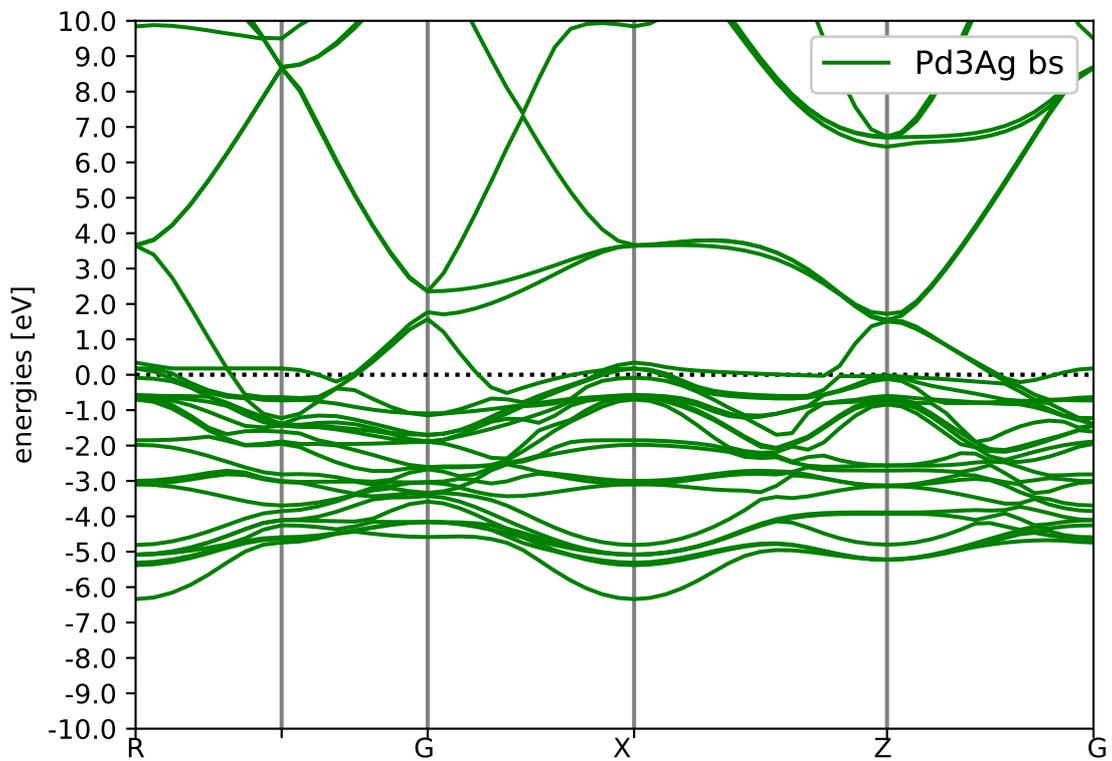}
\end{adjustbox}
\caption{Band structure of bulk~Pd$\rm_{3}$Ag.\label{fig5}}    
\end{figure*}
The\ dos\ peaks\ of\ the valence\ states\ 
show-up\ with\ significant\ amount\ in\ 
the\ energy\ region\ of\ E-E$\rm_{F}$~{$\in$}~[-6,0]~eV,\
see~Fig.~\ref{fig34}.~The\ $d$,~$p$,~$\&$~$s$\ 
orbitals~(Fig~\ref{fig34})\ contribute\ to\ 
the\ occupation\ states\ of\ dos.~The\ 
$d$-orbital\ has\ most\ states\ for\ occupation\ 
followed\ by\ $p$-orbital,~which in turn is\ 
followed by\ $s$-orbital.\ The\ $d$\ states\ 
dominate\ states\ near\ the\ Fermi level,\ 
while\ the\ $p$\ states\ contribute\ in\ the\ 
middle\ energy\ level,\ and\ the\ $s$-states\ 
contribute\ in\ the\ deeper\ energy\ levels.\ 
The\ band\ structure~(Fig.~\ref{fig5})~shows\
no\ energy\ gap,~$\&$~that\ the\ system\
has\ a\ typical\ metallic\ character.The\ dos\ 
plots\ of\ hydrogen,~$\&$~CO~adsorbed\ 
on\ Pd$\rm_{3}$Ag~(111)~is\ presented~in~Fig.~\ref{fig6}.\ 
\begin{figure*}[htbp!]
\centering
\begin{adjustbox}{max size ={\textwidth}{\textheight}}
(\centering a)~{{\includegraphics[scale=0.9]{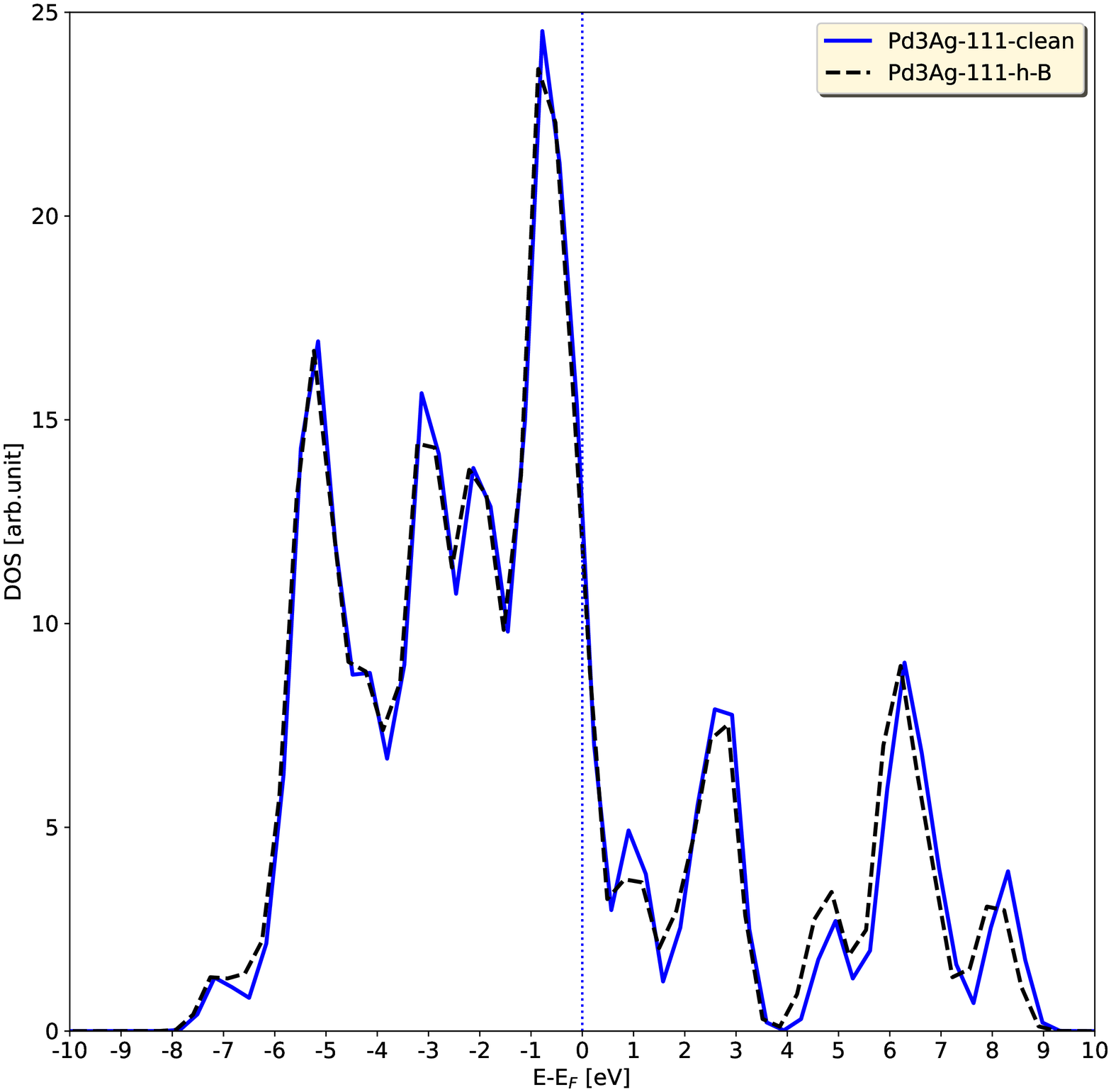}}}
\qquad
(\centering b)~{{\includegraphics[scale=0.9]{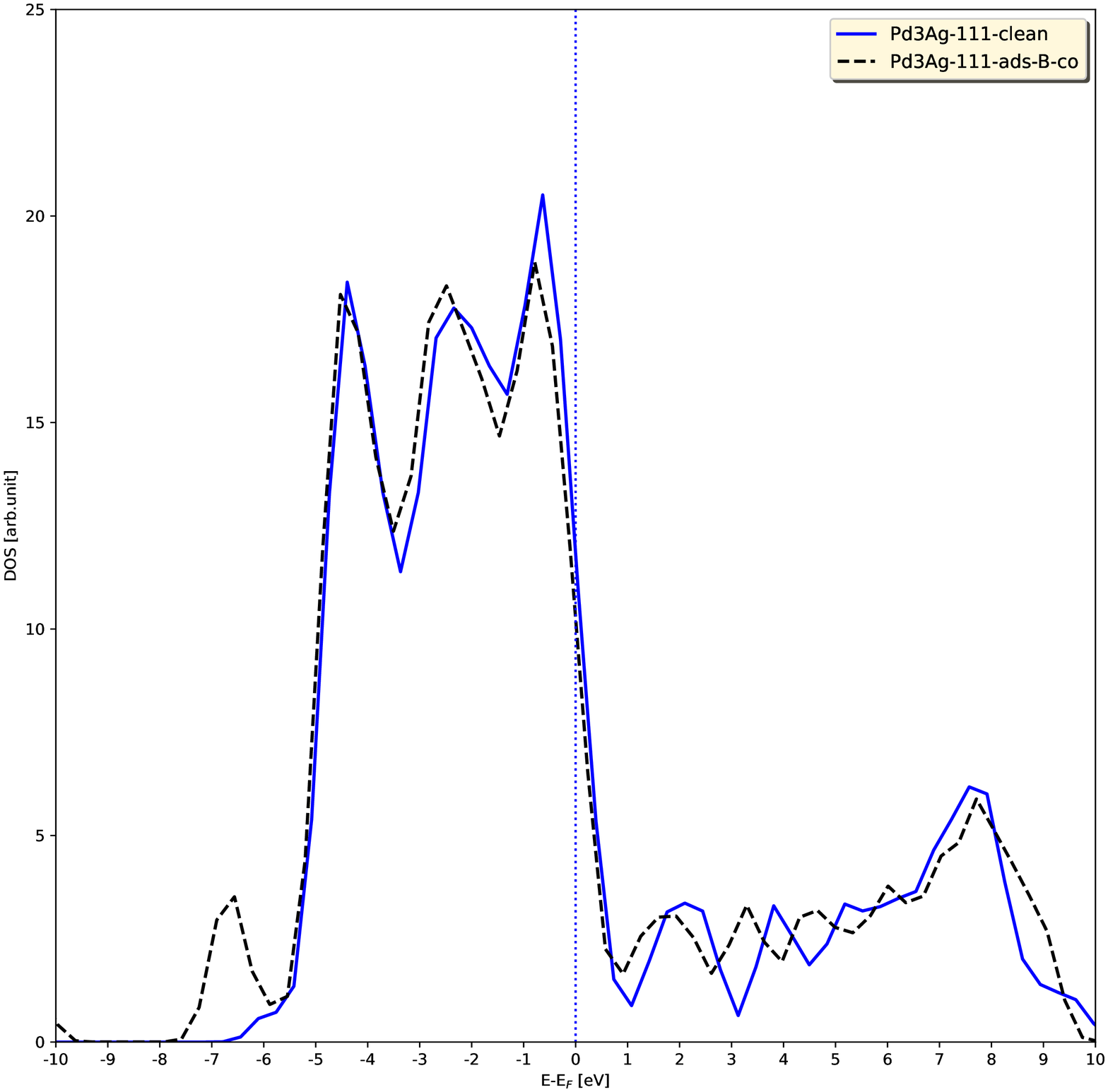}}}
\end{adjustbox}
\caption{(a)~Density of states of Pd$\rm_{3}$Ag~(111)~before\ 
           $\&$~after adsorption\ of~H.~(b)~Density of\ 
           states of Pd$\rm_{3}$Ag~(111)~before~$\&$~after~adsorption\
           of~CO.\label{fig6}}
\end{figure*}
A~slightly\ decreased\ states\ is\ seen\ 
upon\ adsorption,~as shown\ in\ the\ figures,\
due\ to\ partial\ occupation\ of\ the\ empty\ 
orbital\ states\ by\ the\ adsorptions.\ No\ 
major\ energy\ level\ shift\ is\ seen\ between\ 
dos curves\ of\ before\ $\&$~after\ adsorptions.\ 
However,\ the\ nature\ of\ curves\ at\ energy\ 
levels\ E-E$\rm_{F}$~of~-5~eV~$\&$-1~eV\ 
(see\ Fig.~\ref{fig6}~left\ side)\ means\ 
a\ sign\ of\ $d$-$s$\ hybridization\ between\ 
$d$ states\ of\ surface\ atoms~$\&$~$s$ states\ 
of\ the\ adsorbate\ hydrogen.~Similarly,\ from\ 
Fig.~\ref{fig6}~(right\ side),~we\ can\ 
see\ a\ $d$-$p$~hybridization~with\ the\
adsorption\ of\ CO.\ 
\begin{figure*}[htbp!]
\centering
\begin{adjustbox}{max size ={\textwidth}{\textheight}}
(\centering a)~{{\includegraphics[scale=0.5]{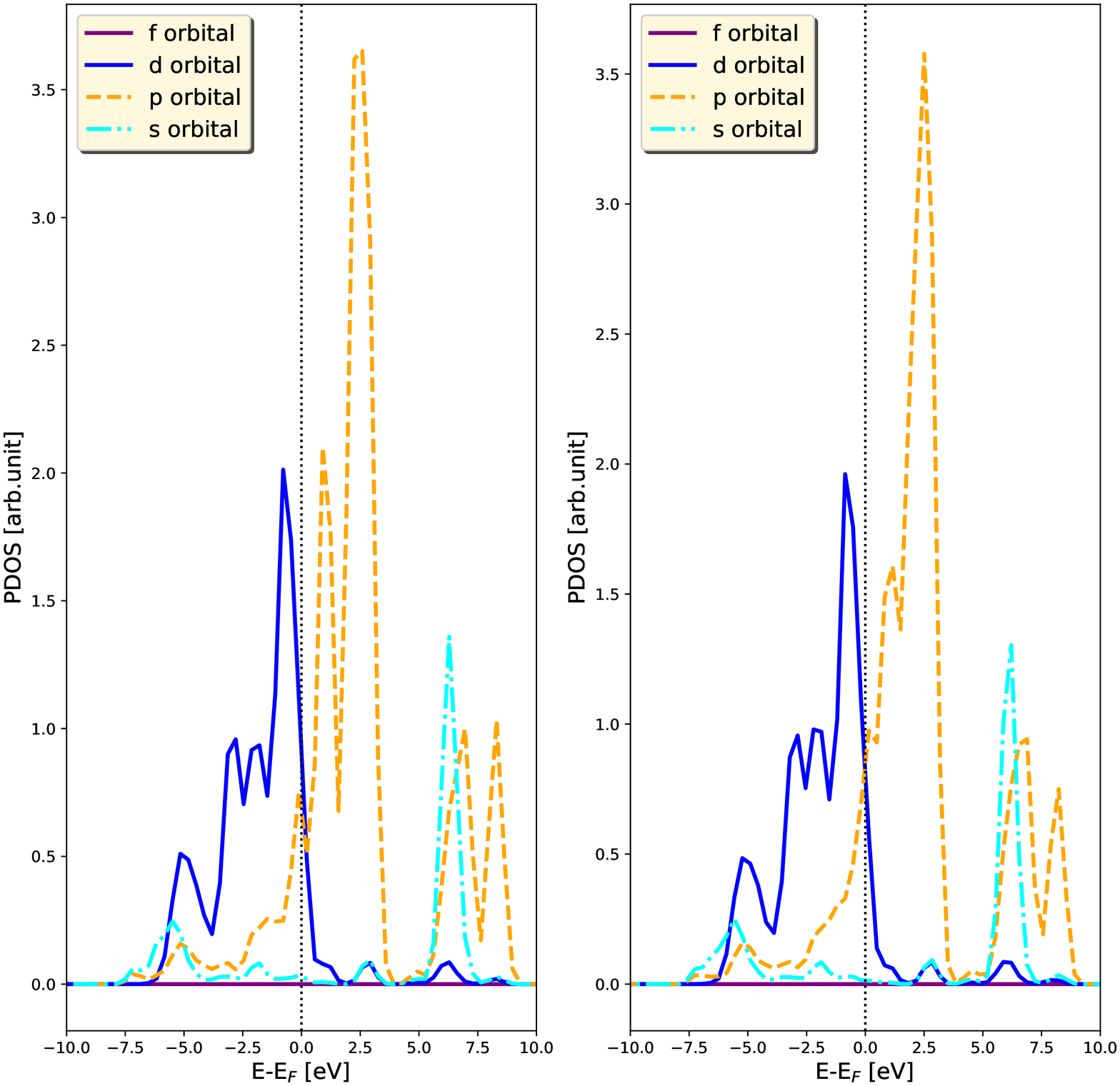}}}
\qquad
(\centering b)~{{\includegraphics[scale=0.5]{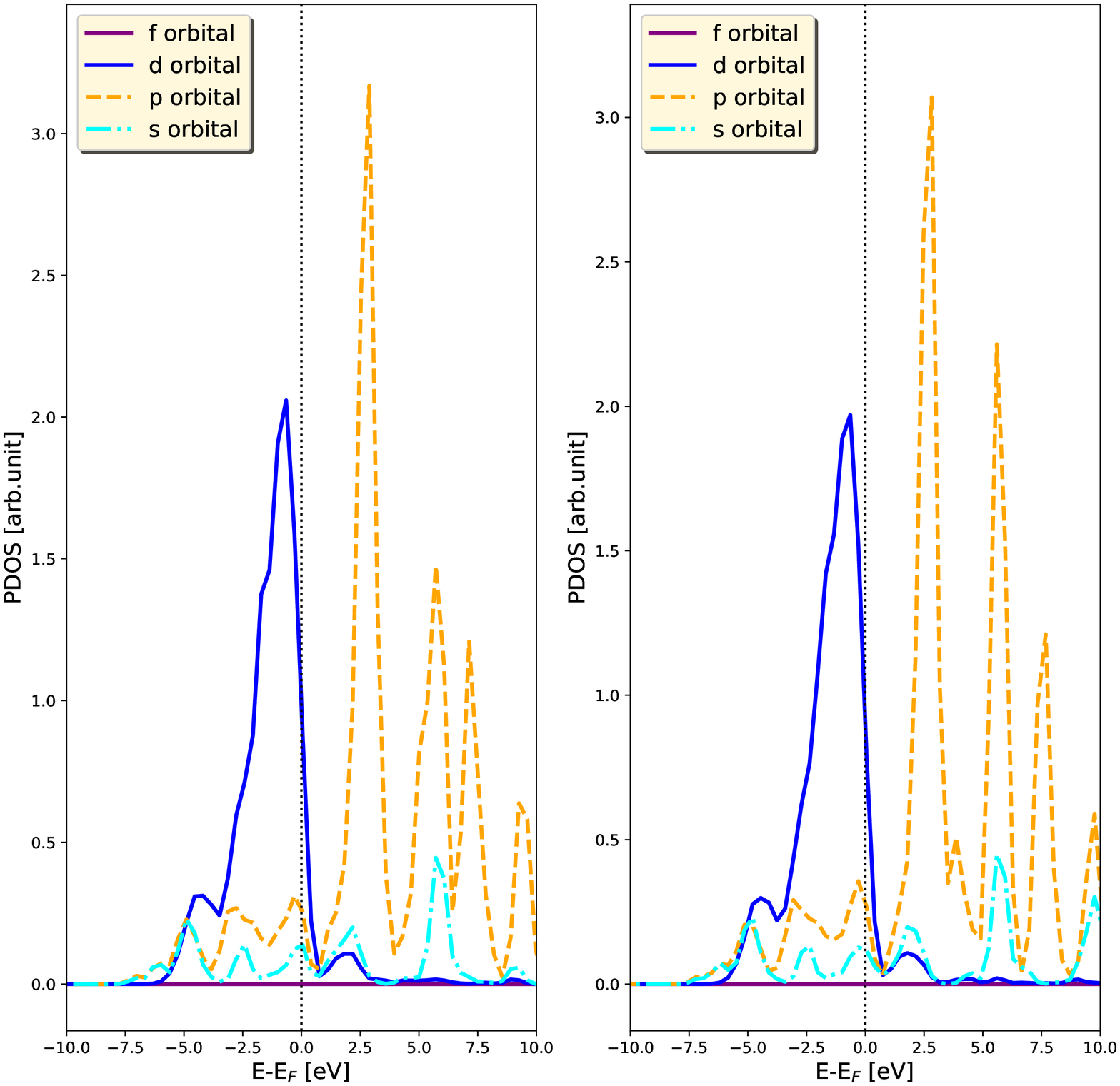}}}
\end{adjustbox}
\caption{(a)~Projected density of states of Pd$\rm_{3}$Ag~(111)~before\
            $\&$~after\ adsorption of~hydrogen.~(b)~Projected\ density\ 
          of states of\ Pd$\rm_{3}$Ag~(111)~before~$\&$~after adsorption\ 
          of~CO.~The\ left\ sides\ within\ each\ graph~(a,~b)~is before\ 
          adsorptions,\ while\ the\ right\ sides\ is\ after\ 
          adsorption.\label{fig7}}
 
\end{figure*}
From\  Fig.~\ref{fig7},\ the\ $d$-$s$,~$\&$\ 
$d$-$p$ orbitals hybridization\ can be\ seen\ 
from\ the\ coinciding\ curve\ peaks~(see the\ 
figure\ on\ left,~$\&$~right\ side,~respectively).\\   

\begin{figure*}[htbp!]
\centering
\begin{adjustbox}{max size ={\textwidth}{\textheight}}
(\centering a)~\includegraphics[scale=0.9]{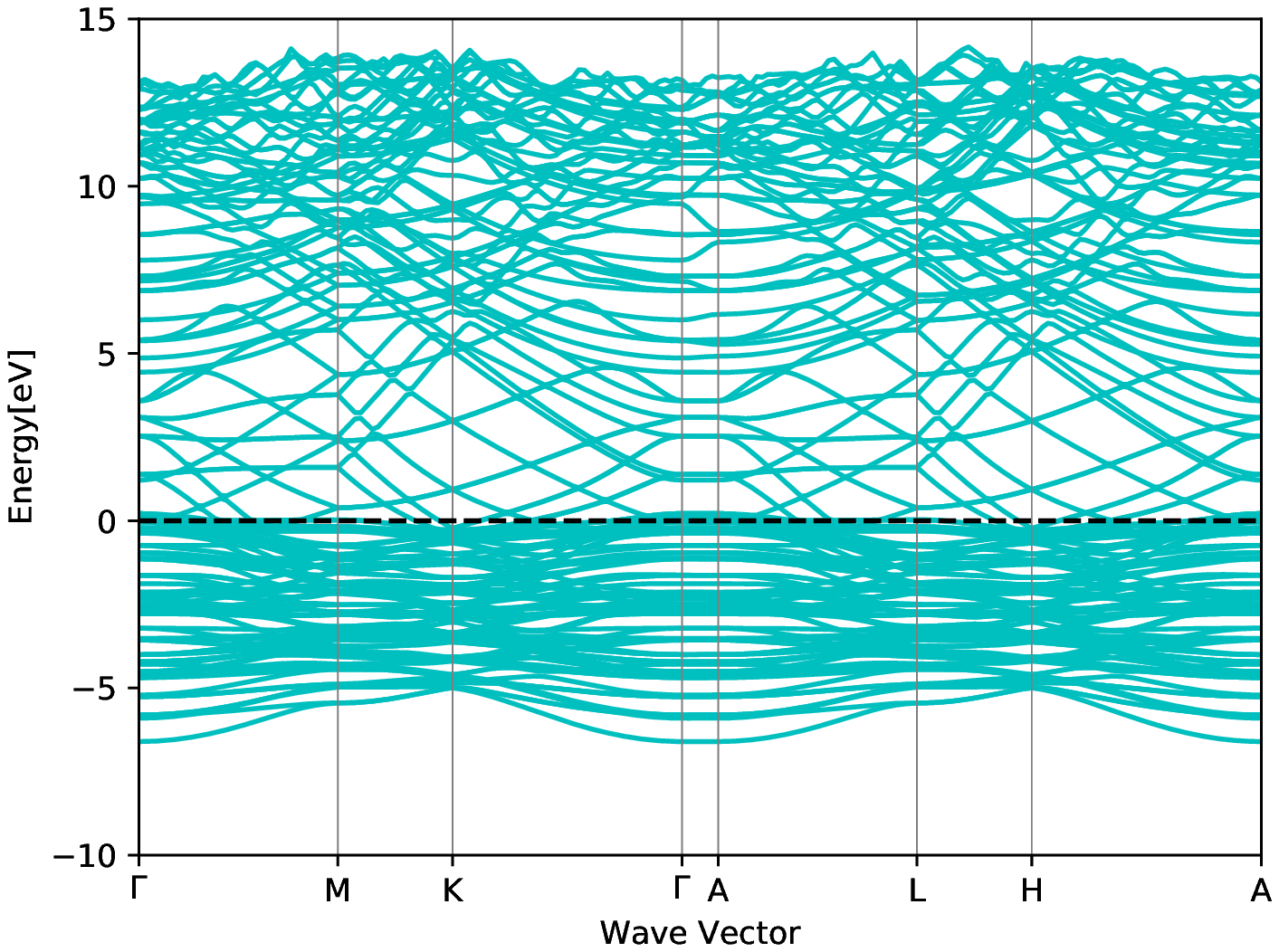}
(\centering b)~\includegraphics[scale=0.9]{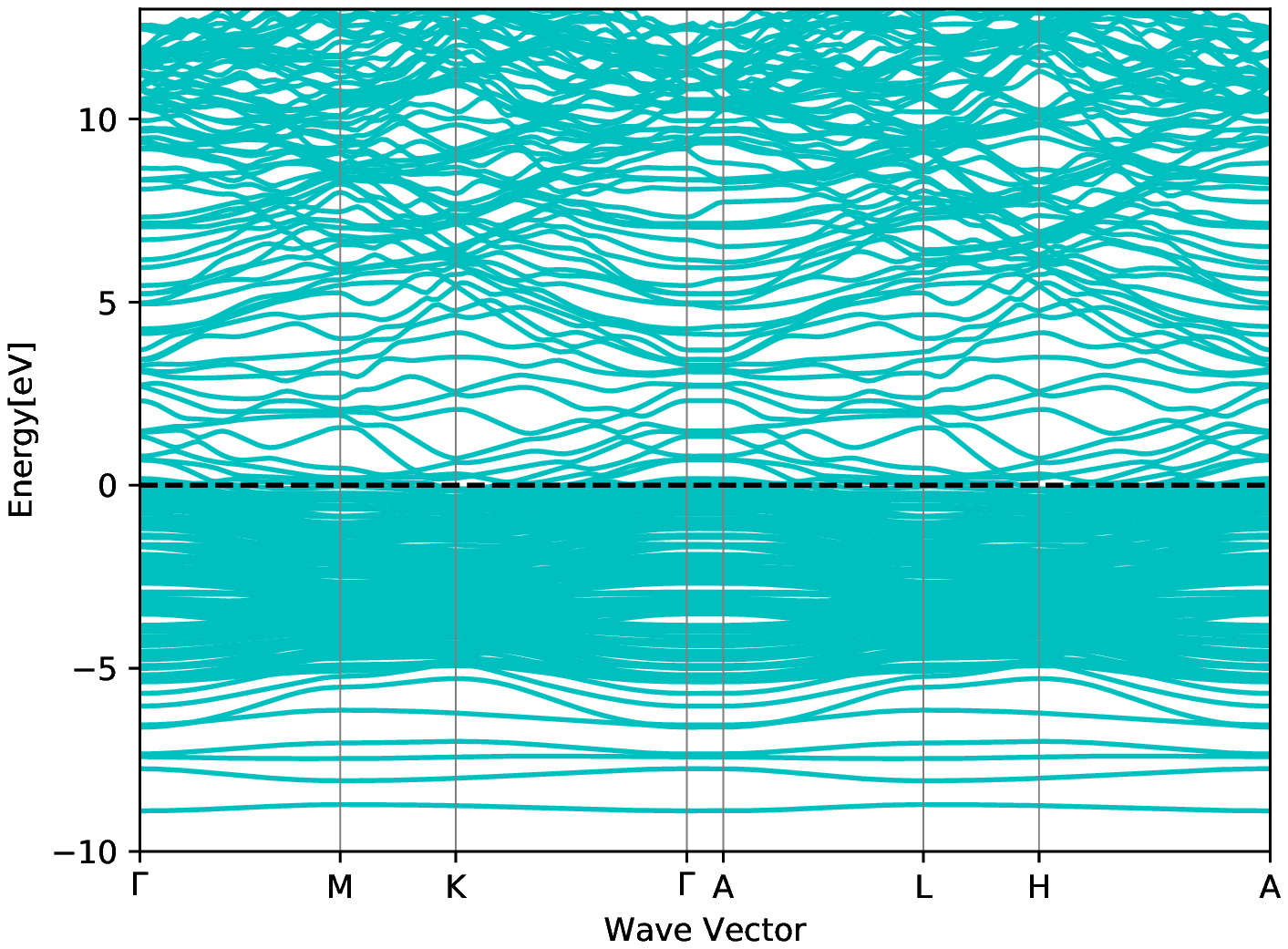}
\end{adjustbox}
\caption{a)~Band structure of clean~Pd$\rm_{3}$Ag~(111)~(left~side).\
         b)~Band structure of Pd$\rm_{3}$Ag~(111)~after~adsorption\ 
         of hydrogen~(right~side).\label{fig8}}
\end{figure*}

By\ the\ adsorptions,\ the\ band\ structure\ 
plots~(Fig.~\ref{fig8})~show\ that\ no\ 
typical\ difference\ to\ the\ electrical\ 
properties\ of\ the\ surfaces\ happens,~except\ 
that\ the\ energy\ curves\ at\ the\ special\ 
k-points\ become\ more\ dense\ due\ to\ an\ 
added\ energy\ eigenvalues\ from\ the\ 
adsorbate\ states.\  

\subsection{KMC\ output\ of\ desorption\ process}
In\ this\ section,~a\ Monte\ Carlo\ simulation\ 
of\ the\ desorption\ process\ of\ adsorbates\ 
from\ the\ alloy\ surfaces\ is\ presented.\ 
In the simulation,~two cases are\ considered,\ 
with\ adsorbate coverage being~$\theta$ = 1 ML~(i.e.,\
1~molecule/atom\ of adsorbate per a ($1\times$1)\ 
surface unit cell).\ The first case\ is by\ 
considering a\ lateral interaction on the\ 
adsorbate which is given by~$15.8~(\theta - 0.25)^{1.3}$~eV,\ 
and the second case is by neglecting lateral\ 
interaction.\
\subsubsection{Simulation of desorption with\ a\ lateral\ interaction}
\begin{figure}[htbp!]
\centering
\begin{adjustbox}{max size ={\textwidth}{\textheight}}
\includegraphics[scale=0.5]{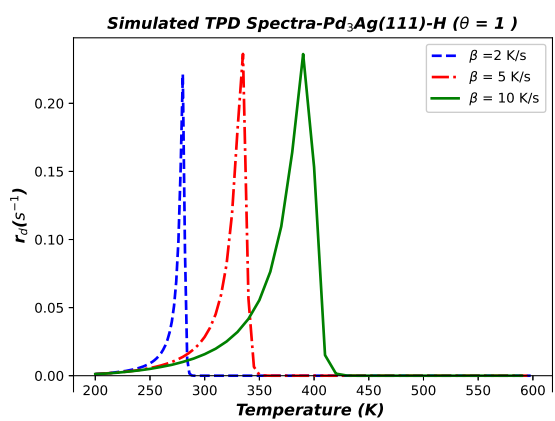}
\end{adjustbox}
\caption{Simulation output\ of desorption of hydrogen 
         from Pd$\rm_{3}$Ag~(111),~within first order\
         for $\beta$ = 2 K/s~(dashed line,~blue color),\
         $\beta$ = 5 K/s~(dash dot line,~red color),\
         and~$\beta$ = 10 K/s~(solid line,~green color),\
         with lateral\ interaction~included.\label{fig9}}
\end{figure}

First,~we present\ the\ desorption\ process\ of\ 
hydrogen adsorbate~as\ follows.\ 
The adsorption energy\ of hydrogen at bridge\ 
site\ of\ Pd$\rm_{3}$Ag~(111)~is 0.34 eV~(Table~\ref{tab3}).\
Figure~\ref{fig9} shows the simulation\ output of\ 
first\ order desorption process by including\ the\
effect\ of\ lateral\ interaction for different\
heating\ rates of\ $\beta$ = 2 K/s, $\beta$ = 5 K/s\ 
and $\beta$ = 10 K/s.\ 
 
\begin{figure}[htbp!]
\centering
\begin{adjustbox}{max size ={\textwidth}{\textheight}}
\includegraphics[scale=0.5]{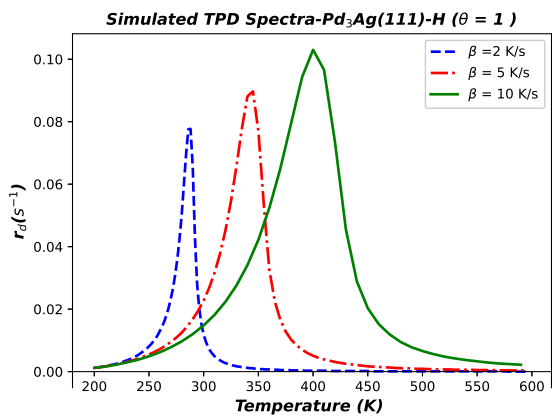}
\end{adjustbox}
\caption{Simulation output\ of desorption of hydrogen 
         from Pd$\rm_{3}$Ag~(111),~within second order\
         for $\beta$ = 2 K/s~(dashed line,~blue color),\
         $\beta$ = 5 K/s~(dash dot line,~red color),\
         and~$\beta$ = 10 K/s~(solid line,~green color),\
         with lateral\ interaction~included.\label{fig10}}
\end{figure}

The desorption peaks\ occur\ at temperatures\ of $T\rm_{m}$ = 280 K\
for~$\beta$ =  2 K/s,~ $T\rm_{m}$ = 335 K~for $\beta$ = 5 K/s,\ 
and $T\rm_{m}$ = 390 K~for~$\theta$ = 10 K/s.~The peaks\ shift\ 
to the right~(higher temperature) as\ the heating rate\ 
increases.\ For\ a simulation\ within\ second order,\ the\ 
output\ is\ presented\ in~Fig.~\ref{fig10}.\

From the graph,\ the\ desorption\ peaks\ occur\ at\
temperatures\ of\ $T_{m}$ = 287 K~for~$\beta$ =  2 K/s),\ 
$T\rm_{m}$ = 345 K~for $\beta$ = 5 K/s,~and $T\rm_{m}$ = 400 K\ 
for $\theta$ = 10 K/s.~The peaks\ shift\ to the right~(higher\ 
temperature) as\ the heating rate\ increases.~The peaks\
occur at a\ relatively higher temperature,~by\ up to 10~K,\
compared\ to the case\ of\ first\ order.\\  

\begin{figure}[htbp!]
\centering
\begin{adjustbox}{max size ={\textwidth}{\textheight}}
\includegraphics[scale=0.5]{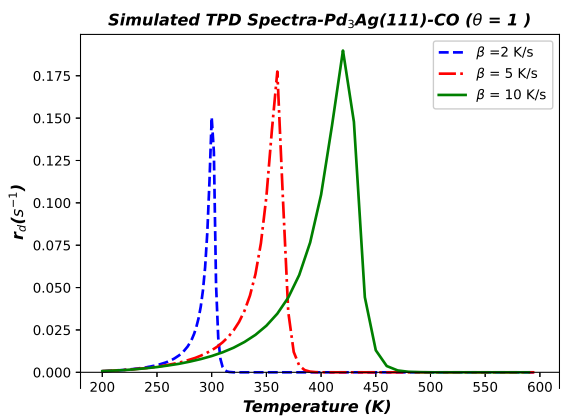}
\end{adjustbox}
\caption{Simulation output\ of desorption of CO\ 
         from Pd$\rm_{3}$Ag~(111),~within first order\
         for $\beta$ = 2 K/s~(dashed line,~blue color),\
         $\beta$ = 5 K/s~(dash dot line,~red color),\
         and~$\beta$ = 10 K/s~(solid line,~green color),\
         with lateral\ interaction~included.\label{fig11}}

\end{figure}

\begin{figure}[htbp!]
\centering
\begin{adjustbox}{max size ={\textwidth}{\textheight}}
\includegraphics[scale=0.5]{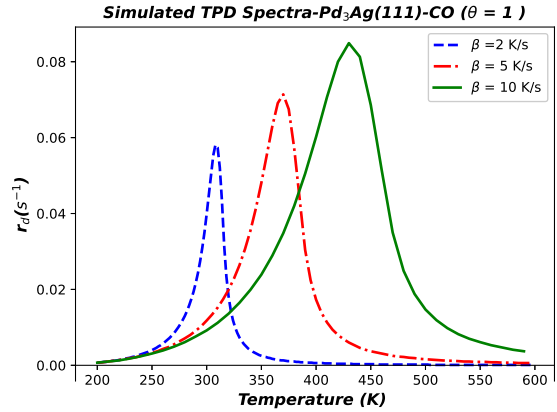}
\end{adjustbox}
\caption{Simulation output\ of desorption of CO\ 
         from Pd$\rm_{3}$Ag~(111),~within second order\
         for $\beta$ = 2 K/s~(dashed line,~blue color),\
         $\beta$ = 5 K/s~(dash dot line,~red color),\
         and~$\beta$ = 10 K/s~(solid line,~green color),\
         with lateral\ interaction~included.\label{fig12}}
\end{figure}

Secondly,\ the simulation\ of desorption rate for\ 
the adsorbate\ CO\ is given\ as\ follows.\  
The adsorption energy of CO at bridge\ site\ of\ 
Pd$\rm_{3}$Ag~(111)~is 1.26 eV~(Table~\ref{tab4}).\ 
The\ outcomes\ of\ simulation\ within\ first\
$\&$~second\ order\ rates\ is\ presented\ in\ 
Figs.~\ref{fig11}~$\&$~\ref{fig12}.

With\ lateral interaction included,~$\&$~within\ 
first\ order\ rate\ concept,\ the\ desorption\ 
peaks\ occur\ at\ temperatures\ of $T\rm_{m}$ = 300 K~for\ 
$\beta$ =  2 K/s),~$T\rm_{m}$ = 360 K~for $\beta$ = 5 K/s,\ 
and $T\rm_{m}$ = 420 K~for $\theta$ = 10 K/s~(see~Fig.~\ref{fig11}).\ 
The peaks\ shift\ to the right~(higher temperature)~as\ 
the heating\ rate\ increases.\ Within\ the second order\ 
rate\ concept,~$\&$~the lateral interaction\ included,\
the desorption peaks occur at temperatures of $T\rm_{m}$ = 308 K\ 
for~$\beta$ =  2 K/s,~$T\rm_{m}$ = 370 K~for $\beta$ = 5 K/s,\ 
and\ $T\rm_{m}$ = 430 K~for $\theta$ = 10 K/s~(see~Fig.~\ref{fig12}).\ 
The peaks\ shift to the right~(higher temperature) as the\ 
heating rate\ increases.~Compared\ to\ the\ peaks\ from\ 
first order rate concept,\ the peaks\ within the second\
order rate concept occur at a\ relatively higher~temperatures,\
by\ up to 10~K.~The width\ of\ desorption curves\ increases\ 
for second order\ process rate compared\ to\ that\ of\ the\ 
first\ order. The\ area under the curves\ is\ related the\ 
amount\ of desorbed\ particles.\
\subsubsection{Simulation of desorption\ without\ a\ lateral\ interaction}
\begin{figure}[htbp!]
\centering
\begin{adjustbox}{max size ={\textwidth}{\textheight}}
\includegraphics[scale=0.5]{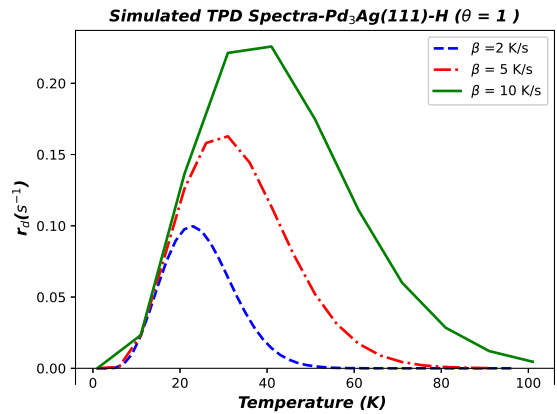}
\end{adjustbox}
\caption{Simulation output\ of desorption of hydrogen\ 
         from Pd$\rm_{3}$Ag~(111),~within first order\
         for $\beta$ = 2 K/s~(dashed line,~blue color),\
         $\beta$ = 5 K/s~(dash dot line,~red color),\
         and~$\beta$ = 10 K/s~(solid line,~green color),\
         without including a lateral interaction~effect.\label{fig13}}
\end{figure}

\begin{figure}[htbp!]
\centering
\begin{adjustbox}{max size ={\textwidth}{\textheight}}
\includegraphics[scale=0.5]{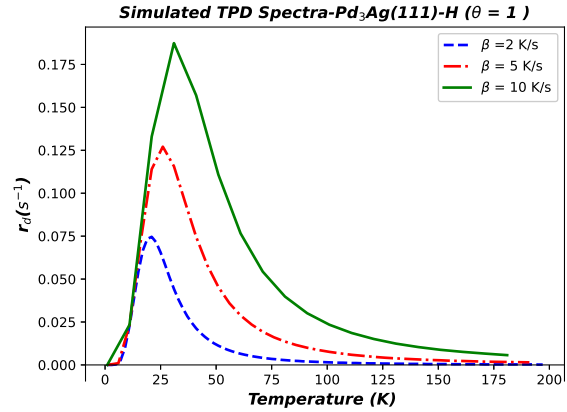}
\end{adjustbox}
\caption{Simulation output\ of desorption of hydrogen\ 
          from Pd$\rm_{3}$Ag~(111),~within second order\
          for $\beta$ = 2 K/s~(dashed line,~blue color),\
          $\beta$ = 5 K/s~(dash dot line,~red color),\
          and~$\beta$ = 10 K/s~(solid line,~green color),\
          without including a lateral interaction~effect.\label{fig14}}
\end{figure}

Wherever\ it\ applies,~1$\rm^{st}$~$\&$~2$\rm^{nd}$\ 
order\ rate\ concepts\ is described\ in\ Eq.~\eqref{rd},\
while\ the\ concept\ of\ lateral\ interaction\ is\
given\ in\ Eq.~\eqref{eqint2}.\ 
In\ this\ case,\ the\ same adsorption energy\ of\ 
hydrogen on\ Pd$\rm_{3}$Ag~(111),\ 0.34~eV,~is\ used,\ 
but\ no\ lateral\ interaction\ is\ used.\ The\ output\ 
of\ the\ corresponding\ simulation\ is\ given in\ 
Figs.~\ref{fig13}~$\&$~\ref{fig14}.\ 

For a first order\ desorption rate\ concept,\ the\ 
desorption peaks\ occur at temperatures of\ 
$T\rm_{m}$ = 21 K~for\ $\beta$ =  2 K/s,\ 
$T\rm_{m}$ = 26 K~for $\beta$ = 5 K/s,~and $T\rm_{m}$ = 31 K\ 
for $\beta$ = 10 K/s~(see Fig.~\ref{fig13}).\ 
The peaks\ shift to the right~(higher\ temperature) as\ 
the heating\ rate increases.\ With\ the\ second\ order\ 
rate\ concept,\ the\ desorption peaks\ occur\ 
temperatures of $T\rm_{m}$ = 23 K\ for $\beta$ =  2 K/s\ 
$T\rm_{m}$ = 31 K~for $\beta$ = 5 K/s,\ 
and $T\rm_{m}$ = 41 K~for $\beta$ = 10 K/s~(see Fig.~\ref{fig14}).\ 
The peaks\ and temperature of the\ maximum desorption\ 
rate shift to the right~(higher temperature)\ as the\ 
heating rate increases.~Similar\ phenomena\ of\ such\ 
shifts\ with\ increased\ heating\ rate\ is\ also\ 
investigated\ for CO desorption from Cu~\cite{zakeri2004monte}.\ 
Furthermore,~the width\ of\ desorption curves\ increases\ 
for second order\ process rate compared\ to\ that\ of\ the\ 
first\ order. 
\subsubsection{Coverage dependence of desorption\ with\ a\ lateral\ interaction}
\begin{figure}[htbp!]
\centering
\begin{adjustbox}{max size ={\textwidth}{\textheight}}
\includegraphics[scale=0.5]{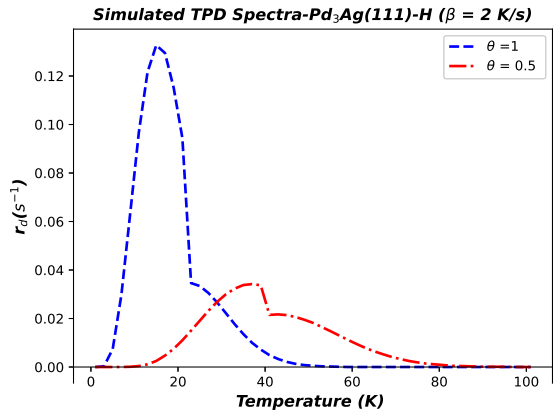}
\end{adjustbox}
\caption{The first order simulation of\ desorption\ 
         hydrogen from Pd$\rm_{3}$Ag~(111) for $\theta$ = 0.5 and 
         $\theta$ = 1 with lateral interaction.
         Color online. Colors:~{$\theta$}=1~(dashed 
         line~$\&$~blue\ color),~{$\theta$}=0.5~(dash-dot
         line~$\&$~red\ color).\label{fig15}}

\end{figure}

In the this section,\ we discuss how\ the peak temperature\ 
depends on\ adsorbate coverage~$\theta$~in the presence\ 
of\ the lateral\ interaction~$\&$~compare it with 
literature.\
With\ hydrogen adsorbate,\ the adsorption energy\ 
on\ Pd$\rm_{3}$Ag~(111)\ is 0.34~eV\ at adsorbate\
coverage of\ 1~ML~(1 molecule of\ adsorbate per a\ 
($1\times$1)~surface unit cell).\ 
At\ adsorbate\ coverage\ of\ 0.5~ML~(i.e.,\ 
corresponding\ to 1 molecule of adsorbate\ per a\ 
($2\times$1)~surface unit cell),\ a\ lateral\ 
interaction of\ -0.14 eV (taken from paper \cite{sraaen}, similar\
work~(adsorption of CO on Pt(111))) is used.~The\ outcome\ 
of\ the\ simulation\ is\ presented\ in\ 
Fig.~\ref{fig15}.\

\begin{figure}[htbp!]
\centering
\begin{adjustbox}{max size ={\textwidth}{\textheight}}
\includegraphics[scale=0.5]{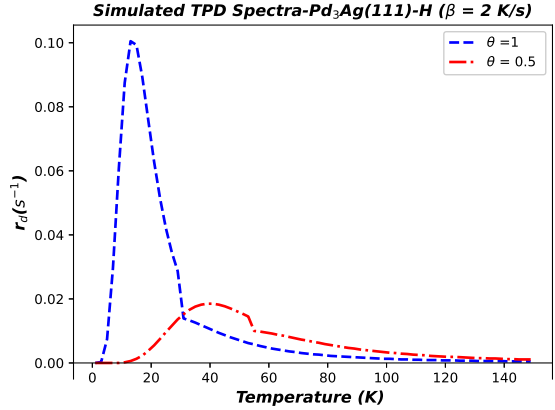}
\end{adjustbox}
\caption{The second order simulation of\ desorption\ 
         hydrogen from Pd$\rm_{3}$Ag~(111) for $\theta$ = 0.5 and 
         $\theta$ = 1 with lateral interaction.
         Color online. Colors:~{$\theta$}=1~(dashed 
         line~$\&$~blue\ color),~{$\theta$}=0.5~(dash-dot
         line~$\&$~red\ color).\label{fig16}}
\end{figure}

From\ the\ Fig.~\ref{fig15},\ it\ can\ be\ seen\ that\ 
the desorption\ peak occurs\ at a\ temperatures of 
$T\rm_{m}$ = 37 K~for~$\theta$ = 0.5 ML, and\ 
$T\rm_{m}$ = 15 K~for $\theta$ = 1 ML.\ The\ temperature\ 
of the peak\ of\ the\ desorption rate increases when\ 
the coverage change from 1~ML to 0.5~ML.\\
For\ a second\ order\ desorption\ rate\ concept,\
the\ outcome\ is\ presented\ in\ Fig.~\ref{fig16}.\
It looks that the\ desorption\ peak\ occurs\ at\ 
temperature of $T\rm_{m}$ = 40 K~for $\theta$ = 0.5 ML,\ 
and $T\rm_{m}$ = 13 K~for $\theta$ = 1 ML.\ The\ 
temperature of the desorption peak increases\ 
when the coverage\ changes from\ 1 ML to 0.5 ML.\\

\begin{figure}[htbp!]
\centering
\begin{adjustbox}{max size ={\textwidth}{\textheight}}
\includegraphics[scale=0.5]{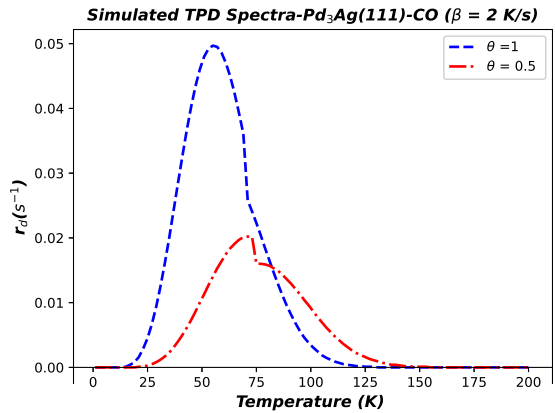}
\end{adjustbox}
\caption{The first order simulation of\ desorption\ 
         CO from Pd$\rm_{3}$Ag~(111) for $\theta$ = 0.5 and 
         $\theta$ = 1 with lateral interaction.
         Color online. Colors:~{$\theta$}=1~(dashed 
         line~$\&$~blue\ color),~{$\theta$}=0.5~(dash-dot
         line~$\&$~red\ color).\label{fig17}}
\end{figure}

\begin{figure}[htbp!]
\centering
\begin{adjustbox}{max size ={\textwidth}{\textheight}}
\includegraphics[scale=0.5]{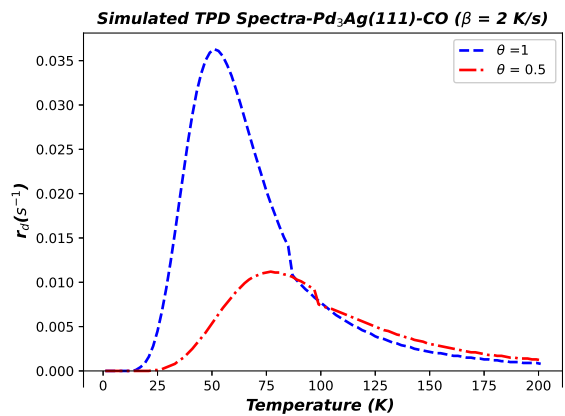}
\end{adjustbox}
\caption{The second order simulation of\ desorption\ 
         CO from Pd$\rm_{3}$Ag~(111)~for~$\theta$ = 0.5 and 
         $\theta$ = 1 with lateral interaction.\
         Color online. Colors:~{$\theta$}=1~(dashed 
         line~$\&$~blue\ color),~{$\theta$}=0.5~(dash-dot
         line~$\&$~red\ color).\label{fig18}}
\end{figure}

With CO adsorbate,~the\ adsorption\ energy of CO\ 
from Pd$_{3}$Ag~(111)~ at 1~ML\ is\ 1.26 eV.\ 
Within first order desorption~rate,~$\&$~using\ 
a\ lateral interaction of -0.14 eV~gives\ 
a\ simulation\ output\ presented\ in\ Fig.~\ref{fig17}.\ 
The\ desorption\ peak\ occurs\ at a\ temperature\
of\ $T\rm_{m}$ = 72 K~for $\theta$ = 0.5 ML,\ 
and $T\rm_{m}$ = 55 K~for $\theta$ = 1 ML.~The\ 
temperature of the desorption\ peak increases\ 
when the coverage changes from 1 ML to 0.5 ML.\
Within\ second order desorption~rate,~the simulation\ 
output\ is presented\ in\ Fig.~\ref{fig18}.\
The\ desorption\ peak\ occurs\ at\ a\ temperature of\
$T\rm_{m}$ = 77 K~for  $\theta$ = 0.5 ML,\ and\ 
$T\rm_{m}$ = 67 K~for $\theta$ = 1 ML.\ The temperature\ 
of the maximum\ desorption peak increases\ when the\ 
coverage change from 1 ML to 0.5 ML.\ There\ is\ 
a confidence\ in such shifts of\ desorption\ 
peaks as coverage decreases,~since\ TPD~experiments\
report similar pattern, albeit on different\ 
system~\cite{sraaen}.\ However,~the\ temperature\ 
values\ could\ be\ possibly\ improved\ further\
by\ computing for exact value of adsorption\ 
energies at coverage of 0.5~ML.~It's value\ 
would\ be\ expected to\ be\ an\ increased\ 
one\ compared\ to the~1.26~eV~value~used\ 
in\ this\ work.\  

\section{Conclusion\label{sec:conc}}
The\ most\ favorable\ adsorptions\ 
have\ adsorption\ energies\ 
in\ the\ range~[0.02, 0.59]\ eV\
for\ a\ hydrogen\ adsorbate\ at\ 
1~ML~adsorbate\ coverage.~With\ 
the\ adsorption\ of\ carbonmonoxide,\
these values are\ in the range\ 
of~[0.10, 1.68]~eV.~The relatively\ 
lower adsorption energies of hydrogen 
compared to CO could indicate that the 
respective ions have better mobility 
within the electrolyte.~Furthermore,\ 
the\ adsorption\ energies\ investigated\ 
for\ CO\ on Pd$\rm_{3}$Ag~surfaces\ are\ 
desorbable\ within\ the\ PEMFC operating\ 
temperature without\ causing\ a\ poisoning\ 
effect.\ Pd$\rm_{3}$Ag\ seems\ to\ be\ 
a\ preferred\ structure,\ compared\ to\ 
PdAg,\ $\&$\ PdAg$\rm_{3}$.\ Furthermore,\
Pd$\rm_{3}$Ag~(111)~is a more\ favorable\ 
as well as reactive surface with the\ 
adsorbates\ at higher fuel supplies,\
thus, it better represents a\ typical\ 
surface of Pd$\rm_{3}$Ag alloy.\ 
A\ charge\ of\ about\ 0.16$e$ per hydrogen\
atom\ appears\ to\ be\ released\ to\ 
the\ system\ for\ electricity\ contribution.\\

The\ only\ possible\ draw\ back of the model\
is\ its\ having\ a slightly\ reduced opportunity\ 
of\ resisting\ extreme external conditions\ due\ 
to pressure\ fluctuations,\ as\ compared\ to\ 
the\ Pt-only\ or\ Pd-only electrode.\ The\ 
adsorbate species\ considered\ in this\ study\ 
can be seen to be\ part of hydrogen\ fuel in\ 
which case 100{$\%$}\ of the hydrogen\ fuel\ 
contains hydrogen species\ or other fuels\ 
such as kerosene,~diesel\ oil,~gasoline,~benzene,\ 
etc,\ which can\ contain up to 10-20$\%$\ carbon\ 
and hydrogen\ components.\ These\ realities\ 
justify\ the\ importance\ of\ dealing\ with\ 
hydrogen and CO\ adsorption,\ and\ desorption\ 
processes.\\

The\ montecarlo simulation\ of\ desorption\ 
process\ suggests that desorption\ peak\ 
of\ carbonmonoxide\ occurs\ at\ 
relatively higher\ temperature~(ca.~400~K)\
compared\ to\ desorption peak\ of\ 
hydrogen which occurs\ at 207~K.\ 
This indicates\ a\ relatively\ 
better operation\ efficiency\ at low\ 
temperatures\ for a hydrogen fuel as\ 
compared to hydrocarbon containing\ 
fuels.~With\ the\ 2$\rm^{nd}$\ order\ 
rate\ concept,\ the\ peak\ of\ the\ 
desorption\ rates\ shifts\ to\ a\ higher\ 
temperature\ compared\ to\ that\ of\ the\
1$\rm^{st}$\ order\ rate.\ Furthermore,\ 
within\ a\ given\ rate\ concept,\ 
the\ temperature\ of\ the\ desorption\ 
peak\ shifts\ to\ a\ higher\ temperature\
as\ the\ adsorbate\ coverage\ decreases\ 
from\ {$\theta$}~=~1~ML~to\ {$\theta$}~=~0.5~ML.\
The\ temperature\ of\ the\ desorption\ 
peaks\ does\ also\ seem\ to\ increase\
when\ heating\ rate\ is\ increased.\
Such\ phenomena\ of\ increases\ in\ 
desorption\ temperatures\ when\ 
a\ heating\ rate\ is\ increased is\ 
also\ investigated\ for CO desorption\ 
from Cu~\cite{zakeri2004monte}.   
Furthermore,\ while\ assuming\ 2$\rm^{nd}$\
order\ rate\ concept,\ the\ width\ of\ 
the\ desorption\ curves\ seem\ to\ be\ 
broadened\ compared\ to\ the\ width\ 
within\ the\ 1$\rm^{st}$\ order\ rate\ 
concept.\ The\ width\ of\ the\ curves\ 
is\ in turn\ directly\ proportional\ 
to\ the\ amount\ of\ desorbed\ particles.\\
In\ recommendation\ for\ a\ need\ area\  
of\ further\ studies,\ the\ authors\ 
commend\ for\ a\ continuous\ study\ 
on\ this\ subject\ which\ focuses\ 
on\ dealing\ with\ transport\ 
processes\ of\ desorbed\ ions\ 
within\ the\ electrolyte\ of\ 
the\ PEMFC.\ 

\section*{Disclosure\ statement}
The authors declare\ that there is no conflict\ 
of interest.

\section*{Acknowledgments}
We are grateful to the Ministry of Education\ 
of\ Ethiopia for financial support.\ The\ 
authors also acknowledge\ the\ Department of\ 
Physics at\ Addis Ababa University and Haramaya\ 
University.\ 
\newpage
\section*{References}
\bibliographystyle{elsarticle-num}
\bibliography{refs.bib}
\end{document}